\pdfoutput=1
\documentclass[a4paper,fleqn,usenatbib]{mnras}
\usepackage{newtxtext,newtxmath}
\usepackage[T1]{fontenc}
\usepackage{ae,aecompl}

\usepackage{graphicx, subfig}	% Including figure files
\usepackage{amsmath}	% Advanced maths commands
\usepackage{amssymb}	% Extra maths symbols
\usepackage[usenames,dvipsnames]{color}
\usepackage{xspace}

\newcommand{\hi}{\ifmmode {\mathrm{H\,\textsc{i}}} \else H\,\textsc{i} \fi}
\newcommand{\hii}{\ifmmode {\mathrm{H\,\textsc{ii}}} \else H\,\textsc{ii} \fi}
\newcommand{\hei}{\ifmmode {\mathrm{He\,\textsc{i}}} \else He\,\textsc{i} \fi}
\newcommand{\heii}{\ifmmode {\mathrm{He\,\textsc{ii}}} \else He\,\textsc{ii} \fi}
\newcommand{\heiii}{\ifmmode {\mathrm{He\,\textsc{iii}}} \else He\,\textsc{iii} \fi}
\newcommand{\Msun}{\ifmmode {\rm M}_{\odot} \else ${\rm M}_\odot$\xspace\fi}
\newcommand{\Mvir}{\ifmmode {M_{\rm vir}} \else $M_{\rm vir}$\xspace\fi}
\newcommand{\Mstar}{\ifmmode {M_\star} \else $M_{\star}$\xspace\fi}
\newcommand{\fesc}{\ifmmode {f_{\rm esc}} \else $f_{\rm esc}$\xspace\fi}
\newcommand{\muv}{\ifmmode {M_{1500}} \else $M_{1500}$\xspace\fi}
\newcommand{\ramsesrt}{\textsc{Ramses-RT}\xspace}

\newcommand{\angstrom}{\text{\normalfont\AA}}

\title[Fluctuating escape fraction of high-$z$ galaxies]{Fluctuating feedback-regulated escape fraction of ionizing radiation in low-mass, high-redshift galaxies}
\author[M. Trebitsch et al.]{Maxime Trebitsch$^{1,2}$\thanks{E-mail:
    trebitsc@iap.fr}, J{\'e}r{\'e}my Blaizot$^{2}$, Joakim Rosdahl$^{2,3}$, Julien Devriendt$^{2,4}$ \newauthor and Adrianne Slyz$^{4}$\\
  {$^{1}$Sorbonne Universit{\'e}s, UPMC Univ Paris 6 et CNRS, UMR 7095,}\\
  {\phantom{$^{1}$}Institut d'Astrophysique de Paris, 98 bis bd Arago, 75014 Paris, France}\\
  {$^{2}$Univ Lyon, Univ Lyon1, ENS de Lyon, CNRS, Centre de Recherche Astrophysique de Lyon UMR5574,}\\
  {\phantom{$^{2}$}F-69230, Saint-Genis-Laval, France}\\
  {$^{3}$\,Leiden Observatory, Leiden University, P.O. Box 9513, 2300 RA, Leiden, The Netherlands}\\
  $^{4}$Sub-department of Astrophysics, University of Oxford, Keble Road, Oxford OX1 3RH, UK}

\date{Accepted 2017 April 28. Received 2017 April 14; in original form 2017 February 13}

\pubyear{2017}

\begin{document}
\label{firstpage}
\pagerange{\pageref{firstpage}--\pageref{lastpage}}
\maketitle


\begin{abstract}
Low mass galaxies are thought to provide the bulk of the ionizing radiation necessary to reionize the Universe. The amount of photons escaping the galaxies is poorly constrained theoretically, and difficult to measure observationally. Yet it is an essential parameter of reionization models.
We study in detail how ionizing radiation can leak from high redshift galaxies. For this purpose, we use a series of high resolution radiation hydrodynamics simulations, zooming on three dwarf galaxies in a cosmological context.
We find that the energy and momentum input from the supernova explosions has a pivotal role in regulating the escape fraction, by disrupting dense star forming clumps, and clearing sight lines in the halo. In the absence of supernovae, photons are absorbed very locally, within the birth clouds of massive stars.
We follow the time evolution of the escape fraction, and find that it can vary by more than six orders of magnitude. This explains the large scatter in the value of the escape fraction found by previous studies.
This fast variability also impacts the observability of the sources of reionization: a survey even as deep as $\muv = -14$ would miss about half of the underlying population of Lyman-continuum emitters.
\end{abstract}


%%%%%%%%%%%%%%%%%%%%%%%%%%%%%%%%%%%%%%%%%%%%%%%%%%

%%%%%%%%%%%%%%%%% BODY OF PAPER %%%%%%%%%%%%%%%%%%

\section{Introduction}
\label{sec:intro}

% EoR
The Epoch of Reionization (EoR) is a transition era in the history of the Universe during which the first structures formed. Until the formation of the first stars and galaxies at redshift $z \sim 15-20$, all the gas in the Universe was neutral. The appearance of these first objects marks the end of the so-called `Dark Ages', and as the first luminous sources formed, they started to emit ionizing radiation, creating ionized \hii regions around them. As the first galaxies assembled, these bubbles gradually grew and merged, until the Universe was fully ionized at $z \sim 6$. The presence of large volumes of neutral hydrogen in the high redshift Universe has initially emerged from the observations of the Gunn-Peterson trough \citep{Gunn1965} in the spectra of high redshift quasars \citep{Becker2001, Fan2001, Fan2006}. It has later been confirmed through the measurement of the Thomson scattering optical depth of Cosmic Microwave Background (CMB) photons on free electrons in the intergalactic medium (IGM) with satellites like the \emph{Wilkinson Microwave Anisotropy Probe} \citep[WMAP, ][]{Hinshaw2013} and \emph{Planck}. The latest results of the \emph{Planck} mission report a Thomson optical depth of $\tau =  0.066 \pm 0.012$, equivalent to a redshift of instantaneous reionization $z_{\rm re} = 8.8^{+1.2}_{-1.1}$ \citep{Planck2016}.

% Sources
The nature of the sources responsible for the production of ionizing ultra-violet (UV) photons necessary to reionize of the Universe is still subject to debate. While quasars are extremely bright, they are likely too rare at $z \sim 6$ to account for all the ionizing budget of reionization \citep[e.g.][]{Willott2010,Fontanot2012, Grissom2014, Haardt2015}. The simplest alternative is the stellar reionization scenario (see for instance \citealt{Kuhlen2012}, or the review by \citealt{Barkana2001}), which postulates that high-redshift star-forming galaxies are accountable for the production of the bulk of the ionizing photons. With the advent of extremely deep surveys such as the Hubble Ultra Deep Field (UDF) survey \citep{Koekemoer2013}, the total production of ionizing photons from star-forming galaxies can be constrained by observations. Based on such campaigns, \citet{Robertson2013} showed that it is necessary to extrapolate the UV luminosity function (LF) of galaxies down to very faint magnitudes of typically $\textrm{M}_{\rm UV} \lesssim -13$ \citep[see also][]{Kuhlen2012, Finkelstein2012} to simultaneously match constraints on the reionization history and the Thomson optical depth. While the recent results of the \emph{Planck} mission reduced the tension between these two probes, and thus the need for early reionization, the  contribution of a yet undetected population of star-forming galaxies to the ionizing budget of the Universe at $z \geq 6$ might still be major if not dominant, unless the fraction \fesc of ionizing photons escaping the galaxies is very high. Similarly, using semi-analytical modelling, \citet{Mutch2016} found that galaxies residing in haloes of mass $\Mvir \sim 10^8 - 10^9 \Msun$ are dominant contributors of the ionizing budget of the Universe before reionization is complete.
This point is strengthened by the recent observational constraints on the faint end of the UV LF at high redshift \citep[see for example][]{Atek2015}, showing no cut-off down to magnitude $\textrm{M}_{\rm UV} \lesssim -15$, consistent with what had been found previously at lower redshift \citep{Alavi2014}. Several studies showed that these galaxies have bluer UV-continuum slopes than their more massive counterparts \citep{Bouwens2009, Bouwens2012, Bouwens2014}, with no significant redshift evolution. This indicates that there is a large reservoir of faint galaxies that must be accounted for when computing the cosmic ionizing budget \citep[see also][]{Bouwens2016}.

% Escape fraction: why
There is a wealth of reionization models, which either use an analytical framework driven by observations of high-$z$ galaxies \citep{Madau1999, Haardt2012, Robertson2013, Robertson2015, Duncan2015}, are semi-analytical \citep[e.g][]{Benson2006, Wyithe2007, Dayal2008, Mitra2015, Mutch2016} or are based on large scale simulations \citep[e.g.][]{Gnedin2014, Ricotti2002, Iliev2006, Iliev2014, Ocvirk2016}, and all of them use \fesc directly or indirectly, the fraction of ionizing photons that escape the galaxy to ionize the IGM, as a key parameter.
In order to quantify the contribution of low mass galaxies, one needs to assess and understand the value of \fesc, which is one of the main sources of uncertainty in the modelling of the reionization history.

% Escape fraction: observations
Direct observations of Lyman continuum (LyC) photons leaking from galaxies have turned out to be extremely challenging. Assessing the escape fraction from individual galaxies is even more difficult, and most studies approach the problem via stacking. In the Local Universe, extremely few LyC leaking galaxies have been found, all exhibiting low escape fractions $\fesc \sim 1 - 10\%$ \citep[][see also \citet{Heckman2011}]{Bergvall2006, Leitet2013, Izotov2016}, and for some other candidates, only upper limits could be secured \citep{Heckman2001}. \citet{Borthakur2014} recently reported the finding of a discovery with $\fesc \sim 20\%$, but taking dust effects into account reduced this to $\fesc \sim 1\%$, further indicating that all low-$z$ LyC leakers have low \fesc. At intermediate redshifts, most efforts have given null results or relatively loose upper limits \citep[e.g.][]{Malkan2003, Siana2007, Siana2010, Cowie2009}, but all report a low upper limit on \fesc. At higher $z$, analysis of individual detections or stacked observational samples shows in general a higher \fesc \citep{Shapley2006, Iwata2009, Vanzella2010, Vanzella2015, Nestor2013, Mostardi2013, Mostardi2015}, sometimes above 50\% \citep{Vanzella2016}, even though some studies found a much lower value for the typical \fesc \citep{Boutsia2011, Grazian2016}. The discrepancy might be due to the high contamination rate of the high-$z$ samples \citep{Siana2015}, and to the very uncertain properties of dust at these redshifts. \citet{Bergvall2013} suggested that selection effects might bias the low-$z$ quest for LyC leaking galaxies.

% Escape fraction: theory
In parallel to the observational attempts to determine the escape fractions of star-forming galaxies, a strong theoretical effort has been made over the past two decades. Analytical estimates or calculations using very idealized geometries have found a fairly low value for \fesc in typical galaxies \citep{Wood2000, Clarke2002, Fernandez2011}, quite sensitive to the gas distribution in the ISM \citep{Ciardi2002}.  More recently, several simulations have been undertaken, either in an idealized galaxy setup or in a full cosmological context \citep[e.g.][]{Wise2009, Wise2014, Razoumov2010, Kim2013, Paardekooper2013, Kimm2014, Yajima2014, Ma2015, Kimm2017}. These numerical experiments can give important insights into the escape of radiation in a more realistic context. However, these simulations yield vastly different results: for instance, while \citet{Gnedin2008} and \citet{Wise2009} found that \fesc increases with the mass of the halo hosting the galaxy, other studies such as \citet{Yajima2011}, \citet{Kimm2014}, \citet{Wise2014} or \citet{Paardekooper2015} found the opposite trend, and \citet{Ma2015} found no trend at all. In practice, it is very difficult to compare directly the results from these studies. The halo masses span over six orders of magnitude, and the numerical methods employed are tremendously different: Smooth Particle Hydrodynamics vs. grid codes, coupled radiative transfer vs. post-processing, various degrees of numerical resolutions, etc. Maybe even more decisively, all simulations modelling galaxies in a cosmological context must rely on sub-resolution recipes to describe star formation and the various associated feedback processes. It is extremely hard to compare directly the results of these different simulations, since even the detailed modelling of a single physical process will strongly affect the outcome of a simulation (see for instance \citet{Kimm2015} for a comparison of various implementations of supernova feedback). Furthermore, for a given simulation, there is an intrinsic scatter in the value of \fesc: \citet{Kimm2014} found that there is a large spread in the \fesc vs. halo mass relationship, and \citet{Paardekooper2015} showed that the escape fraction seems to be affected by a large variety of physical parameters.

% Scope of the paper
Using a sample of three galaxies simulated with very high resolution and state of the art subgrid models, we propose to go a step forward in studying the detailed mechanism leading to the escape of ionizing radiation, with the ultimate goal of understanding the scatter in \fesc.
The combination of our subgrid models for star formation and feedback has so far only been employed in simulations of mini-haloes, which have been found to be of minor importance for reionization \citep{Kimm2017}. Our study extends the mass ranged probed by this model to haloes of masses around a few $10^9\Msun$. Our new model for star formation, which causes a very bursty and clustered star formation, also extends the work of \citet{Kimm2014}, which focuses on a mass range similar to the one presented in this study.
% Organization
This paper is organized as follows: in Sect.~\ref{sec:simulations}, we describe our simulation methodology and our sub-resolution recipes. In Sect.~\ref{sec:basic}, we present general properties of our simulated sample. We study in Sect.~\ref{sec:fesc} the evolution of the ionizing escape fraction at the halo scale, and in Sect.~\ref{sec:escape}, we discuss the mechanisms that controls the escape from the galaxy. We discuss possible improvements of this work in Sect.~\ref{sec:discussion}, and summarize our results in Sect.~\ref{sec:summary}.

\section{Description of the suite of simulations}
\label{sec:simulations}

We introduce here the suite of simulations that analyse use through this work.
We focus on a selection of three haloes of masses up to $3\times 10^{9}\Msun$ that we follow down to $z \simeq 5.7$, when the Universe was $\sim$ 1 Gyr old.

\subsection{RHD simulations with \ramsesrt}
\label{sec:ramsesrt}

We run cosmological simulations with \ramsesrt \citep{Rosdahl2013, Rosdahl2015}, a public multi-group radiative transfer (RT) extension of the adaptive mesh refinement (AMR) code \textsc{Ramses}\footnote{\label{fn:ramses}\url{https://bitbucket.org/rteyssie/ramses/}} \citep{Teyssier2002}. The evolution of the collisionless dark matter (DM) and stellar particles is followed using a particle-mesh solver with cloud-in-cell interpolation. The evolution of the gas is followed by solving the Euler equations using a second-order Godunov scheme. We use the HLLC Riemann solver \citep{Toro1994}, with a MinMod total variation diminishing scheme to reconstruct the intercell fluxes. For all the simulations, a Courant factor of 0.8 has been used. We use a standard quasi-Lagrangian refinement strategy, in which a cell is refined if $\rho_{\rm DM} \Delta x^3 + \frac{\Omega_{\rm DM}}{\Omega_b}\rho_{\rm gas} \Delta x^3+ \frac{\Omega_{\rm DM}}{\Omega_b} \rho_* \Delta x^3 > 8\ m_{\rm DM}^{\rm HR}$, where $\rho_{\rm DM}$, $\rho_{\rm gas}$ and $\rho_*$ are respectively the DM, gas and stellar densities in the cell, $\Delta x$ is the cell size, and $m_{\rm DM}^{\rm HR}$ is the mass of the highest resolution DM particle. In a DM-only run, this would allow refinement as soon as there are at least 8 high-resolution DM particles in a cell.

The RT module propagates the radiation emitted by stars in three groups (describing the average \hi, \hei and \heii ionizing photons) on the AMR grid using a first-order Godunov method with the M1 closure for the Eddington tensor. We highlight that this moment-based method can handle an arbitrary number of ionizing sources. However, because the radiation is evolved with the same timestep as the gas, we use the reduced speed of light approximation, with a reduced speed of light of $\tilde{c} = 0.01 c$, where $c$ is the real speed of light.
The ionizing photons are produced by star particles at each timestep using the \textsc{Galaxev} model of \citet{Bruzual2003}, assuming a \citet{Chabrier2003} initial mass function (IMF). Note that while most of the radiation is emitted by stars younger than 5 Myr, we continue to take into account the photons emitted by older star particles. For this set of simulations, we use the on-the-spot approximation, assuming that any ionizing photon emitted by recombination will be absorbed locally. We hence ignore straight-to-ground level recombination and the associated emission of ionizing radiation from the gas.
The coupling with the hydrodynamical evolution of the gas is done by incorporating the local radiation while computing the non-equilibrium thermochemistry for the hydrogen and helium.

\subsection{Initial conditions}
\label{sec:IC}

We assume a flat $\Lambda$CDM cosmology consistent with the \emph{Planck} results \citep[$\Omega_\Lambda = 0.6825, \Omega_m = 0.3175, h = 0.6711$ and $\Omega_b = 0.049$,][]{Planck2013}. We select three haloes from a $10\, h^{-1}$ comoving Mpc DM-only simulation with $512^3$ particles ($m_{\mathrm{DM}} \simeq 6.6\times 10^5\ \Msun$) initialized at $z = 100$. The haloes are selected in the final output at $z \sim 5.7$ with \textsc{HaloMaker} \citep{Tweed2009}. They all are relatively isolated (there is no halo with more than half its mass within more than 10 virial radii), with quiet merger histories and masses of approximately $10^8, 10^9$ and $3\times 10^9\, \Msun$ at $z = 5.7$.

The haloes are then resimulated for 1 Gyr at much higher resolution using the `zoom-in' technique. The initial conditions for the resimulations are produced using the \textsc{Music}\footnote{\label{fn:music}\url{https://bitbucket.org/ohahn/music/}} code \citep{Hahn2011}. For each resimulation, we refine the mass distribution such that the DM particle mass is $m_{\mathrm{DM}} \sim 1.9 \times 10^3 \Msun$ in the zoomed-in region, equivalent to a $4096^3$ particle simulation. This corresponds to an AMR coarse grid level of $l = 12$ in the zoomed-in region, and we allow for 9 more levels of refinement, giving a most refined cell size of $\Delta x = 10h^{-1}\textrm{Mpc} / 2^{21} = 7.1$ pc.

Since we do not track the formation of population III (pop III) stars and early metal enrichment, we assume an initial gas phase metallicity $Z = 10^{-3} Z_\odot = 2 \times 10^{-5}$ everywhere, consistent with metal enrichment from primordial mini-haloes \citep{Whalen2008}.

\subsection{Physical modelling}
\label{sec:subgrid}

\subsubsection{Star formation}
\label{sec:SF}

Because of the huge dynamical range involved in galaxy and star formation, we cannot afford to resolve all the stages of the gravitational collapse from the cosmological scale down to the formation of individual stars. Following what is usually done in galaxy formation studies, we use a subgrid recipe to model the collapse of a gas cloud below the resolution limit. In this section, we briefly describe the main ideas behind our star formation model. The reader interested in the detailed implementation is referred to Devriendt et al. (in prep, see also \citealt{Kimm2017}) for a complete discussion of the model.

The main physical ingredient of this model is the idea that at the star-forming cloud scale, turbulence in the ISM can act as an effective extra pressure support. We thus define a `turbulent Jeans length' as
\begin{equation}
  \label{eq:lamjt}
  \lambda_{J, \mathrm{turb}} = \frac{\pi \sigma_{\rm gas}^2 + \sqrt{36\pi c_{\rm s}^2 G \Delta x^2 \rho_{\rm gas} + \pi^2 \sigma_{\rm gas}^4}}{6 \pi G \rho_{\rm gas} \Delta x},
\end{equation}
with $\sigma_{\rm gas}$ the velocity dispersion of the gas computed using the velocity gradients around the cell, $\rho_{\rm gas}$ is the gas density and $c_{\rm s}$ is the local sound velocity. We identify star forming sites as cells with $4 \Delta x \geq \lambda_{J, \mathrm{turb}}$, i.e. cells which are unstable even with this additional turbulent support.

For these star forming cells, we use an approach similar to that of \citet{Rasera2006} to convert $\epsilon \rho_{\rm gas} \Delta x^3 \Delta t/t_{\rm ff}$ of gas into star particles during one timestep $\Delta t$, where $\epsilon$ is the local star formation efficiency and $t_{\rm ff} = \sqrt{3\pi / 32 G \rho_{\rm gas}}$ is the gas free fall time. The primary difference with the recipe of \citet{Rasera2006} is that the star formation efficiency is not a constant, but rather derived from the local properties of the gas \citep{Hennebelle2011, Federrath2012}:
\begin{equation}
  \label{eq:sfr_ff}
  \epsilon \propto e^{\frac{3}{8}\sigma_s^2}\left(1 + \mathrm{erf}\left(\frac{\sigma_s^2 - s_{\rm crit}}{\sqrt{2\sigma_s^2}}\right)\right),
\end{equation}
where $\sigma_s  = \sigma_s(\sigma_{\rm gas}, c_{\rm s})$ characterizes the turbulent density fluctuations, $s_{\rm crit} \equiv \mathrm{ln}\left(\frac{\rho_{\rm gas, crit}}{\rho_0}\right)$ is the critical density above which the gas will be accreted onto stars (where $\rho_0$ is the mean density of the cloud), and $\rho_{\rm gas, crit} \propto (\sigma_{\rm gas}^2 + c_{\rm s}^2) \frac{\sigma_{\rm gas}^2}{c_{\rm s}^2}$. The exact formulae are given in \citet{Federrath2012}, using their multi-scale model based on \citet{Padoan2011}, coined `multi-ff PN' model.
As a result, instead of being smoothly distributed in the ISM, the star forming gas will be gathered in clumps, leading to a very clustered star formation process. We adopt a minimum stellar particle mass of $m_\star \simeq~135 \Msun$, leading to the explosion of at least one supernova per particle assuming a \citet{Chabrier2003} IMF. On average, for the most massive galaxy in our sample, the mean mass for the star particles is around $460 \Msun$.

\subsubsection{Feedback from massive stars}
\label{sec:FB}

Our simulations implement both radiative feedback from stars and type II supernova feedback. For the radiative feedback, we only take photoionization heating into account, since it is believed to be the dominant channel of radiative feedback \citep{Rosdahl2015a}. More specifically, star particles emit ionizing radiation as described in \ref{sec:ramsesrt}, which will inject energy and heat the surrounding medium.

We use the recent mechanical supernova (SN) feedback implementation described in \citep[][see also \citealt{Rosdahl2017}]{Kimm2014, Kimm2015}, with one instantaneous supernova event per star particle after a 10 Myr delay, consistent with the typical lifetime of a SN\textsc{ii} progenitor of $15 \Msun$. The main motivation for this model is to describe correctly the momentum transfer from the SN at all stages of the Sedov blast wave, from the early adiabatic expansion to the late snowplough phase. In practice, the amount of momentum deposited in the surrounding cells depends on the properties (density and metallicity) of the gas in these cells. Note that because the gas distribution can be anisotropic around a star particle, the SN explosion can in principle result in an anisotropic energy deposition as well. A feature of our new star formation model is that the stars will form in a very clustered fashion, and as a result the total amount of energy deposited by the supernovae from a group of star particles is very similar to what would happen with more massive star particles.

\subsubsection{Gas cooling and heating}
\label{sec:cooling}

\ramsesrt features non-equilibrium cooling by explicitly tracking the ionization state of 5 species (H, H$^+$, He, He$^+$, He$^{++}$). The primordial cooling rate is directly computed from these abundances. We include an extra cooling term for metals, using tabulated rates computed with \textsc{Cloudy} \citep[last described in][]{Ferland2013} above $10^4$~K. We also account for energy losses via metal line cooling below $10^4$~K following the prescription of \citet{Rosen1995}. We approximate the effect of the metallicity by scaling linearly the metal cooling enhancement. Note that we do not take into account the impact of the local ionizing flux on the metal cooling. 
We use an homogeneous metallicity floor of $Z = 10^{-3} Z_\odot$ in the whole box to account for the lack of molecular cooling and to allow the gas to cool down below $10^4$~K. We ensured after the fact that our choice of initial metallicity leads to a redshift of first star formation in the progenitors of our main haloes roughly similar to that found by studies of minihaloes including H$_2$ cooling \citep[e.g.][]{Kimm2017}.
Since we do not include any meta-galactic UV background in our simulation either, we only take into account the local photoheating rate. While this will not be valid at the end of the Epoch of Reionization, when the ionized bubbles are overlapping, this holds for our isolated galaxies that are likely to reionize themselves.

\subsection{Computing the escape fraction}
\label{sec:fescmethod}

A last methodological point to discuss is the way we measure the instantaneous escape fraction. With \ramsesrt, we have access to the local photon flux anywhere at any time. We can then simply compute the escape fraction as the total flux crossing a given boundary divided by the intrinsic luminosity of the sources. Unless stated otherwise, we always compute the escape fraction at the virial radius.

\begin{figure*}
  \centering
  \includegraphics[width=\linewidth]{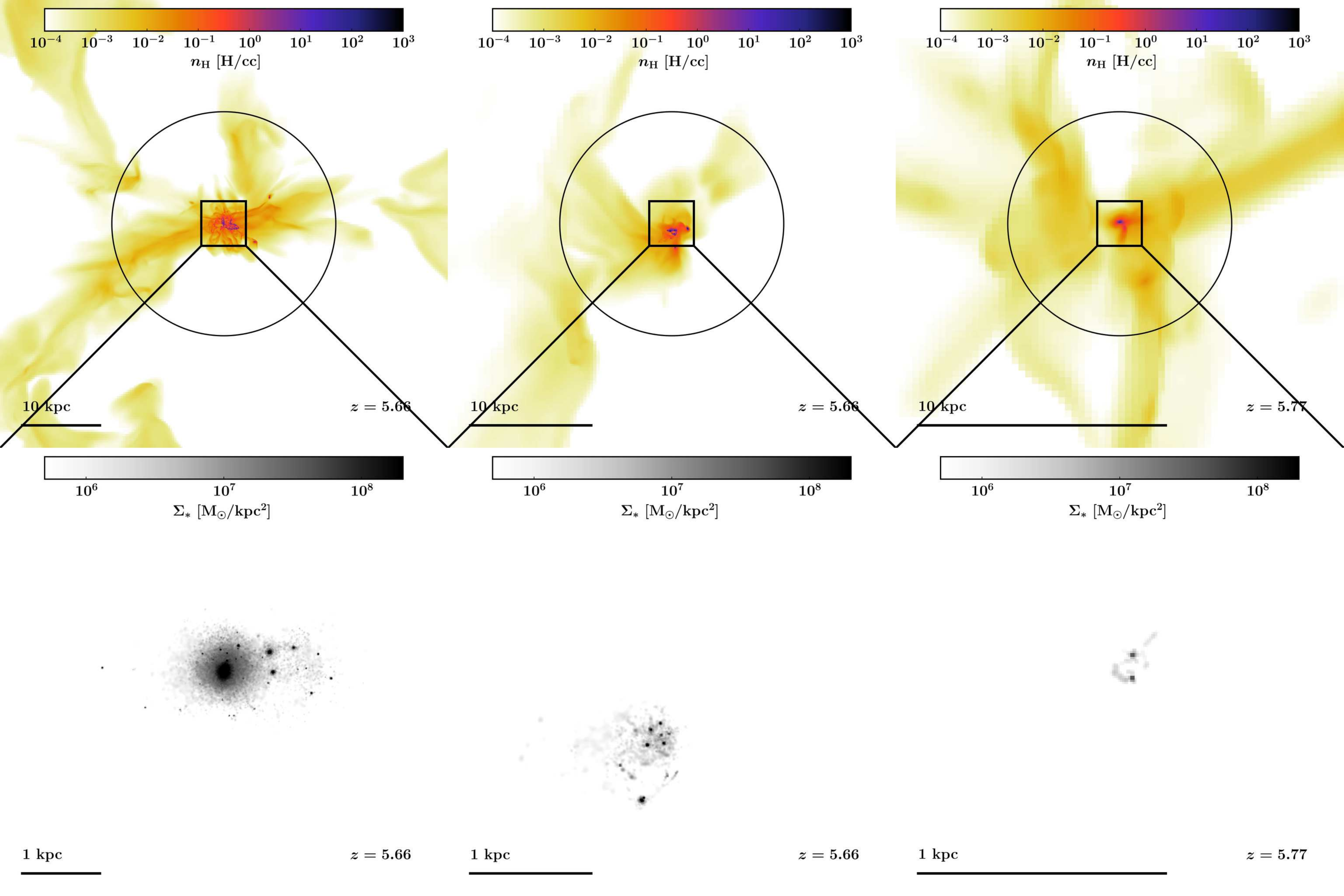}
  \caption{Mass-weighted projection of the gas density ({\em top row}) and zoomed stellar surface density ({\em bottom row}), as indicated by the black boxes, for the three galaxies at the end of each simulation, ordered by decreasing mass from left to right: $3\times 10^9\, \Msun, 10^9\, \Msun$ and $10^8\, \Msun$. The black circle shows the virial radius of each halo, while the physical scale and the redshift of the snapshot are indicated in the bottom left and right corners, respectively.}
  \label{fig:maps}
\end{figure*}

However, since the speed of light is finite (and even considerably reduced in our simulations), it takes time for photons to propagate from their source to the point at which we compute the escape fraction. For a steady source of photons, this delay would be unimportant, but as we will see in Sec.~\ref{sec:basic}, the star formation is very bursty in our simulations. This means that the galaxy will flicker. A photon crosses 10 kpc in approximately 30 kyr at the full speed of light. At our reduced speed of light, this timescale is increased by a factor 100, meaning that in our simulation, the radiation emitted in the centre of a halo with $R_{\rm vir} \sim 10$ kpc would only reach the virial radius after 3 Myr, which is comparable to the lifetime of massive stars, and of the same order of magnitude as the duration of the bursts. It is therefore necessary to take this delay into account when we compute the escape fraction.

We compute the angle averaged escape fraction at a distance $r$ from the centre of the halo at any time $t$ as
\begin{equation}
  \label{eq:fesc}
  \fesc(r, t) = \frac{\int\bmath{F_{\rm out}}(t) \cdot \bmath{\hat{r}} \ \mathrm{d}\Omega}{\sum_i m_*^i \dot{n}_{\rm ion}^i(t - r/\tilde{c})},
\end{equation}
where $\bmath{F_{\rm out}}$ is the outgoing flux of photons, $\bmath{\hat{r}}$ is the local radial direction from the centre of the halo, $m_*^i$ is the mass of a star particle in unit mass, and $\dot{n}_{\rm ion}^i$ is the ionizing photon production rate per unit mass for a simple stellar population of age $t$. We take $\bmath{F_{\rm out}} = \bmath{F_{\rm ion}}$ if the local flux of photons $\bmath{F_{\rm ion}}$ is indeed going outward and 0 otherwise, which prevents us from computing negative contributions to the escape fraction coming e.g. from a satellite or a star-forming clump infalling onto the main galaxy. This is making the assumption that all the photon sources are located at the centre of the halo, leading to a common time delay for all stars.

While this is not always valid in theory (e.g. a star-forming clump closer than the others to the virial radius), we will check in Sect.~\ref{sec:fescestimate} whether this is a good approximation.
For this purpose, we introduce a second way of estimating the escape fraction, through ray-tracing. We generate $N$ directions using the \textsc{HEALPix} \citep[Hierarchical Equal Area isoLatitude Pixelation,][]{Gorski2005} decomposition of the sphere into $N = 12 \times 4^\ell$ equal-area pixels. We typically use $\ell = 2-3$, or $N = 192-768$ directions, unless stated otherwise. For each direction $j$, we can compute the optical depth $\tau_{\hi}^{i,j} = \sigma_{\hi} N_{\hi}^{i,j}(r)$ at a given distance $r$ for each star particle $i$, where $\sigma_{\hi}$ is the average cross-section for the \hi-ionizing radiation, and $N_{\hi}^{i,j}$ is the \hi column density seen from the star $i$ in the direction $j$.
As $\sigma_{\hi} = 3.3\times 10^{-18}\ \mbox{cm}^2$, any line of sight with $N_{\hi}^{i,j} \geq 3 \times 10^{17}\ \mbox{cm}^{-2}$ would be optically thick to ionizing photons.
The luminosity-weighted angle-averaged escape fraction can then be expressed as
\begin{equation}
  \label{eq:fescray}
  \fesc = \frac{\sum_i L_{\rm ion}^i \bar{T}_i}{\sum_i L_{\rm ion}^i},
\end{equation}
with $\bar{T}_i = \langle e^{-\tau_{\hi}^{i,j}} \rangle_{j}$ the angle-averaged transmission for the $i^{\rm th}$ star particle, and $L_{\rm ion}^i$ its ionizing luminosity. While this does not take the halo light crossing time into account, it is in principle more robust since it does not rely on the assumption that all sources are at the centre of the halo. This can also be employed to estimate the \emph{individual} escape fraction of each star particle. We will discuss in Sect.~\ref{sec:fescestimate} the differences between these two methods on the estimated value of \fesc.

\section{Galaxy properties}
\label{sec:basic}

\begin{figure*}
  \centering
  \includegraphics[width=\linewidth]{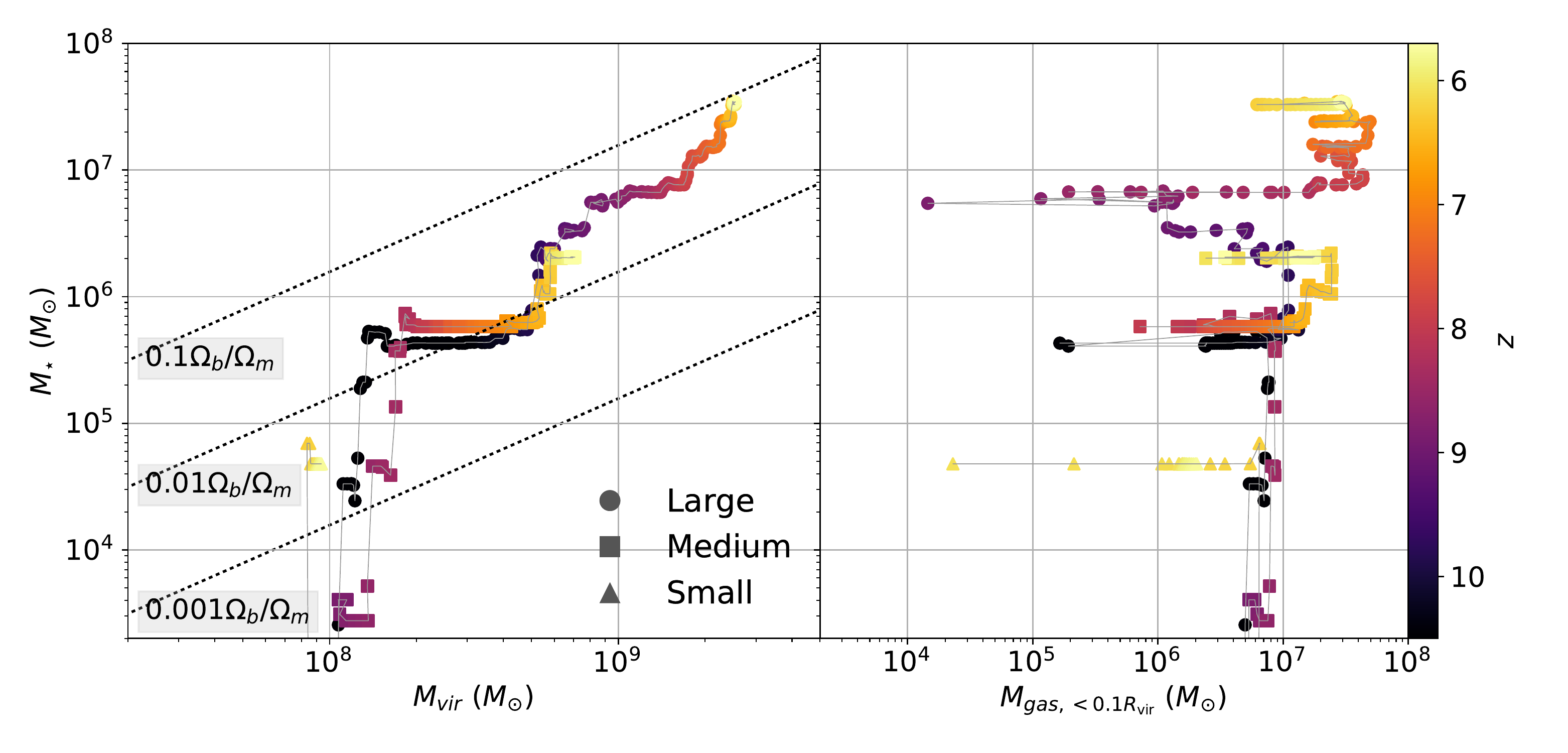}
  \caption{\emph{Left}: Stellar mass - halo mass relationship for the three haloes. Each halo is indicated by a different symbol: filled circles ($3\times 10^9\, \Msun$ halo), squares ($10^9\, \Msun$ halo) and triangles ($10^8\, \Msun$ halo),  the colour of the which indicates the redshift. The three diagonal dotted lines indicate constant baryonic fractions 10\%, 1\% and 0.\% from top to bottom.
    At low mass, episodes of star formation are followed by long periods of quiet accretion. Roughly between 1\% and 10\% of the baryons in each halo is converted into stars.     \emph{Right}: Relationship between stellar mass and gas mass within $0.1 R_{\rm vir}$. After each star formation episode, most of the gas is ejected out of the galaxy.}
  \label{fig:mvirmstar}
\end{figure*}

In Fig.~\ref{fig:maps}, we illustrate  from left to right the gas density and stellar surface density of the three galaxies at the end of each simulation, in decreasing order of halo mass. The stellar distribution is very clumpy and irregular, as expected for such low mass galaxies. While all three haloes seem to be embedded in large scale filaments, we see at least for the first two haloes a disturbed gas morphology, shaped by the strong SN feedback affecting these galaxies.

In Fig.~\ref{fig:mvirmstar}, we show the mass assembly history of each of the three simulated haloes by comparing the stellar mass to the total halo mass (on the left) or to the gas mass within 10\% of the virial radius (on the right). The coloured symbols represent the positions of the haloes on the \Mstar - \Mvir plane at a given redshift. The diagonal dotted lines indicate, from top to bottom, 10\%, 1\% and 0.1\% of the baryonic mass fraction. Overall, in our simulations, roughly between 1\% and 10\% of the baryons are locked in stars at any given time. This is slightly higher than what has been found by previous studies, but given the lack of observational constraints at $z \sim 6-10$ for low-mass galaxies, the large scatter in the \Mstar - \Mvir relation allows for haloes such as the one we present here.
The recent work of \citet[][Fig.~6]{Miller2014} shows that in the Local Volume, low-mass galaxies have higher stellar mass to halo mass ratios than predicted by abundance matching. While these galaxies are not exact analogues of higher $z$ ones, this is consistent with our findings, and despite the large uncertainties, this mitigates the importance of differences between our simulations and other results e.g. from abundance matching techniques.

We note two striking features of Fig.~\ref{fig:mvirmstar}. First, for all three galaxies, the first episodes of star formation occur at $\Mvir \sim 10^8 \Msun$ (or similarly at $M_{\rm gas} \sim 2\times 10^7 \Msun$), independent of the redshift, which is the typical `atomic cooling' limit. This is most likely a resolution issue: we re-ran the lowest mass halo at a higher resolution, and found that the first stars formed earlier.
Second, we see that the stellar mass increases only episodically, especially at low masses. As an example, let us focus on the `large' halo. At $z \sim 12$, a first dramatic star formation event happens, pushing the stellar mass of the galaxy to $\Mstar \simeq 5\times 10^{5}\ \Msun$. This episode is followed by a long plateau during which star formation is shut off as the halo still grows in mass, until $z \sim 9.8$ (approximately 120 Myr later). This discontinuous stellar mass assembly directly results from the supernova feedback, which strongly reduces (or even shuts off) star formation by ejecting most of the gas out of the galaxy, so the galactic gas reservoir has to replenish before fuelling any new star formation (right panel). At the same time, the growth of the halo mass (and also the total gas mass) is virtually unaffected.

\begin{figure}
  \centering
  \includegraphics[width=\linewidth]{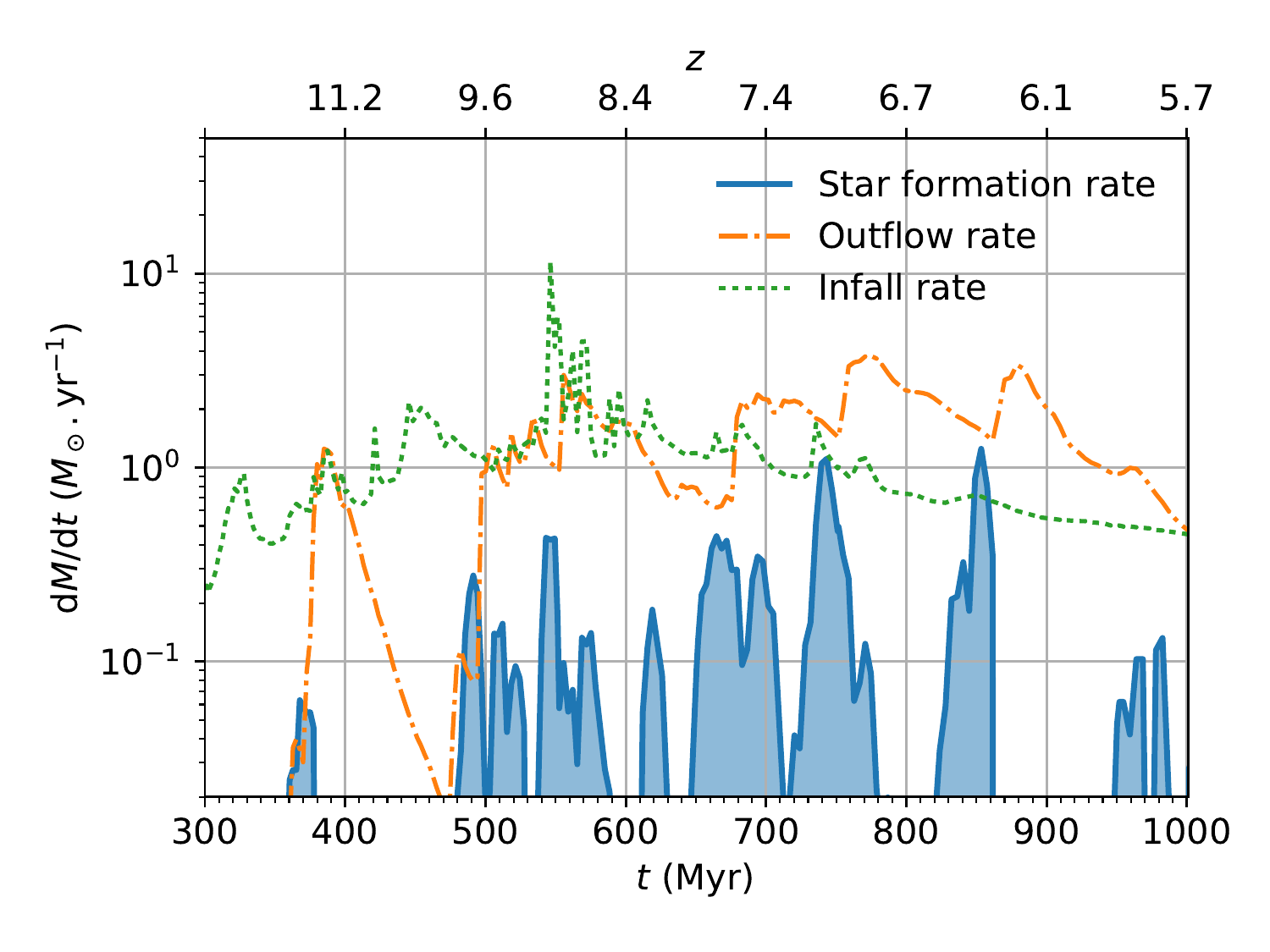}
  \caption{Star formation (blue, solid line), outflow at $R_{\rm vir}$ (orange, dotted line) and infall at $R_{\rm vir}$ (green, dash-dotted line) history of the most massive halo as a function of time. Each star formation episode is followed by a massive outflow event.}
  \label{fig:sfr_outflow}
\end{figure}

The bursty behaviour of the star formation is further illustrated as a blue filled surface in Fig.~\ref{fig:sfr_outflow} for the most massive halo, where we show its evolution as a function of time. Here, we define the star formation rate as the mass of stars newly formed averaged over 10 Myr. We also display the outflow rate at the virial radius with the orange dash-dotted line.
Typically, during a star formation episode, the galaxy can reach a SFR as high $1\ \Msun.\mbox{yr}^{-1}$, and more typically $0.4\ \Msun.\mbox{yr}^{-1}$, each episode lasting for 20 to 50 Myr. This burstiness has been seen in other numerical studies \citep[e.g.][]{Wise2009, Kimm2014, Hopkins2014, Shen2014}, and is regulated by supernovae explosions that heat the gas or even eject it, and delay the formation of new dense, star-forming clumps. Indeed, we see that approximately 10 Myr after the beginning of each star formation episode, the outflow rate rises up to 2 to 4 $\Msun.\mbox{yr}^{-1}$, denoting the presence of strong winds, with mass loading factor above unity. The evolution of the outflow rate with time follows the same highly variable pattern as the SFR: because star formation is shut off, after all the massive stars have ended their lives as supernovae, the galactic winds gradually weaken.

These winds are responsible for the presence of the plateaus in \Mstar in Fig.~\ref{fig:mvirmstar}: the strong outflows quench star formation for some time. For the `medium' halo, this lasts for about 200 Myr, while for the `large' halo, the plateau only lasts for 100 Myr. One reason for this is that the infall rate at the virial radius (the green dotted line in Fig.~\ref{fig:sfr_outflow}) is typically higher by a factor of 2 to 3 for the large halo compared to the medium one. In the large halo, the time needed to refill the gas reservoir after the massive outflow is therefore significantly shorter, and star formation therefore resumes faster.

\section{Halo escape fraction}
\label{sec:fesc}

We now discuss the ionizing properties of the galaxies in our sample with the goal of understanding the large galaxy to galaxy variation in the escape fraction that has been found in other studies.

\subsection{Time evolution of the escape fraction}
\label{sec:fesc_t}

\begin{figure}
  \centering
  \includegraphics[width=\linewidth]{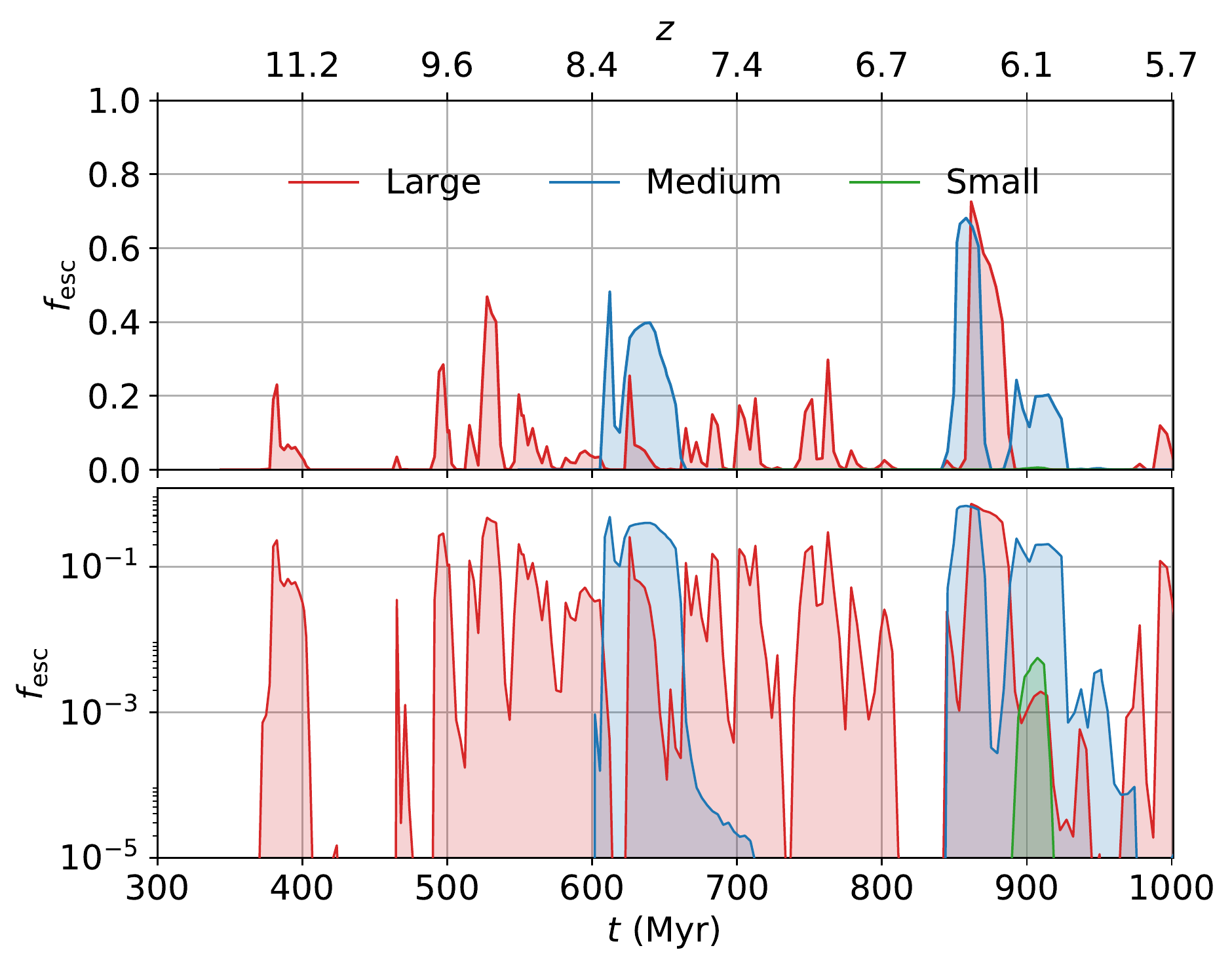}
  \caption{Evolution of the escape fraction of ionizing radiation \fesc with time for all three haloes. The upper panel is in linear scale, while the lower panel is in log scale. All haloes display a very bursty behaviour, \fesc varying from 0 to sometimes more than 60\% in a very short time. The smallest halo only reaches $\fesc \simeq 0.5\%$ during a short burst at $t \simeq 900$ Myr and therefore remains indistinguishable from 0 in the upper panel.}
  \label{fig:fesc}
\end{figure}

In Fig.~\ref{fig:fesc}, we show the evolution of the angle averaged escape fraction of \hi-ionizing radiation for the three haloes, which has a very bursty behaviour. The most massive halo of our study (in red) experiences six episodes during which $\fesc > 20\%$, each of them lasting for 10 to 50 Myr. These episodes are followed by periods of much lower ($\leq 10^{-4}$) escape fraction. The medium halo (in blue) experiences a similar alternation of high and low escape fraction, with \fesc higher than 20\% for long periods. For the smallest halo (in green), there is only one peak of escape fraction, around $z \simeq 6.1$, but the galaxy only reaches a relatively low escape fraction of $\fesc \sim 0.5\%$. This is because the galaxy undergoes only one star formation event. Except for that small halo, our values are broadly consistent with the findings of \citet{Kimm2014} and \citet{Paardekooper2015} who find that for haloes of $\Mvir \sim 10^{8} - 10^9 \Msun$, \fesc is typically between $0.1\%$ and $2\%$, but can be as high as $\sim 100\%$. We must however note that these studies present quantities averaged over an ensemble of galaxies (from a few hundreds for \citealt{Kimm2014} to a few ten thousands for \citealt{Paardekooper2015}), and the intrinsically stochastic nature of the very strong variations in \fesc prevents us from comparing a time-averaged value to the ensemble average of the aforementioned studies. 

\begin{figure}
  \centering
  \includegraphics[width=\linewidth]{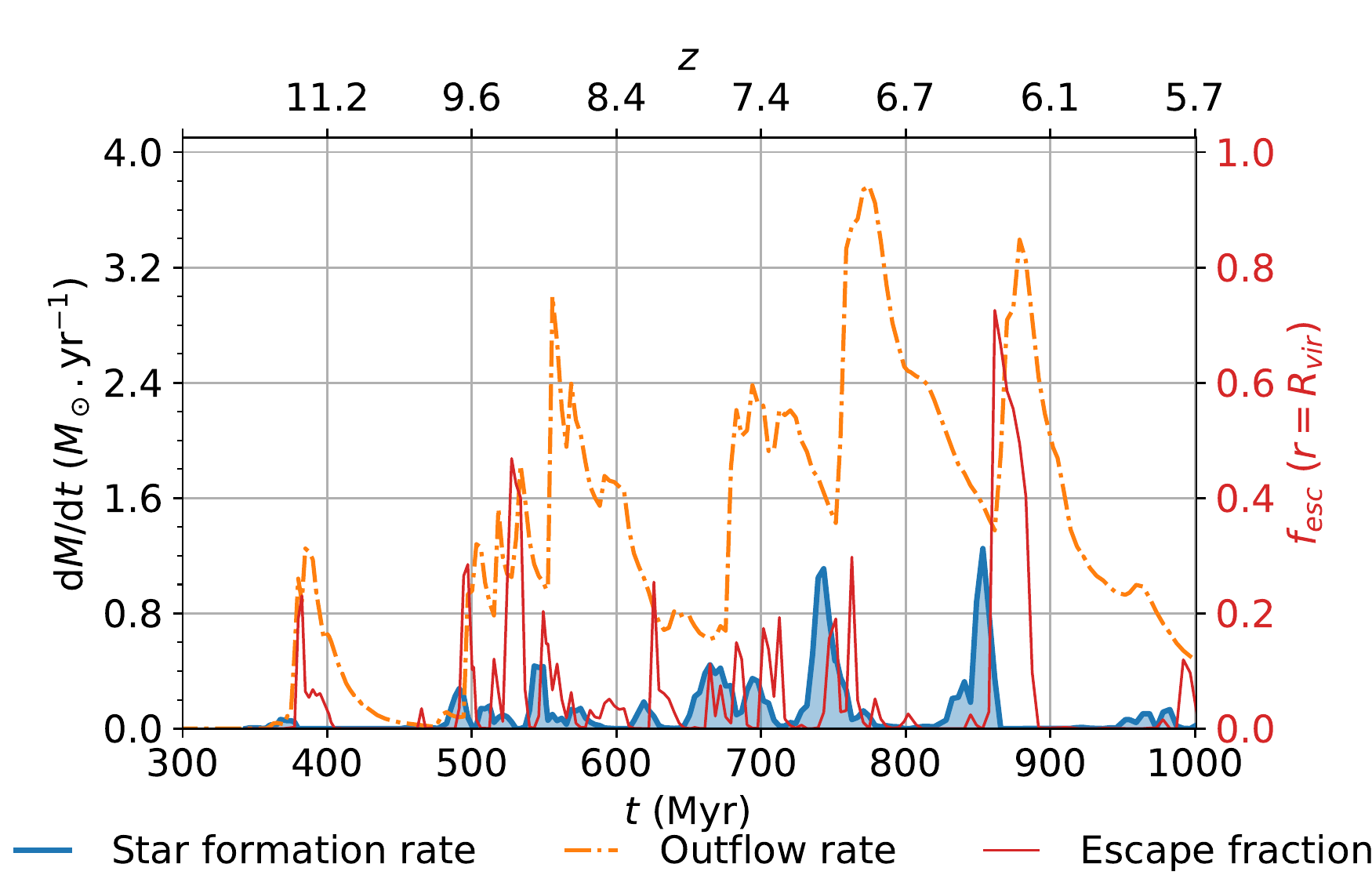}
  \caption{Evolution of the star formation rate (blue), outflow rate at $R_{\rm vir}$ (dash-dotted, orange) and escape fraction (red) for the most massive halo. The escape fraction starts to rise at the same time as the outflow rate, and typically 10 Myr after the beginning of a star formation event.}
  \label{fig:fesc_sfr_out}
\end{figure}

The quick variations over several orders of magnitude of the escape fraction reflect the burstiness of the star formation histories of the three galaxies that we discussed in Sect.~\ref{sec:basic}. We illustrate the correlations between these two quantities by plotting in Fig.~\ref{fig:fesc_sfr_out} the escape fraction (red), the star formation rate (blue) and the outflow rate (dash-dotted orange) for the most massive halo in our sample. Associated with the bursts of star formation, we see that the escape fraction systematically jumps to its highest value right at the onset on the wind. This sharp transition indicates that as soon as there is a hole in the ISM, radiation is able to escape. This typically happens with a 10 Myr delay with respect to the beginning of a new event of star formation, corresponding to the age at which star particles experience the supernova explosion with our modelling (see Sect.~\ref{sec:basic}). While this time delay of $\sim 10$ Myr has been used in other studies \citep[e.g.][]{Kimm2014, Wise2014, Ma2015}, we must tread carefully in that direction, and not take this number at face value: this is the direct result of our subgrid recipe for supernovae, which explode 10 Myr after the birth of the star particle.
We also note that the peak of \fesc is usually reached much more rapidly than the maximum outflow rate: this is because while we compute both quantities at the virial radius, radiation travels this distance much faster than the gas carried in outflows. In addition to limiting the star formation, the supernova explosions completely alter the morphology of the gas in and around the galaxies. The combination of outflows and shock-heating of the gas clears lines of sight around the galaxy and allow ionizing radiation to freely flow into the IGM.

\begin{figure}
  \centering
  \includegraphics[width=\linewidth]{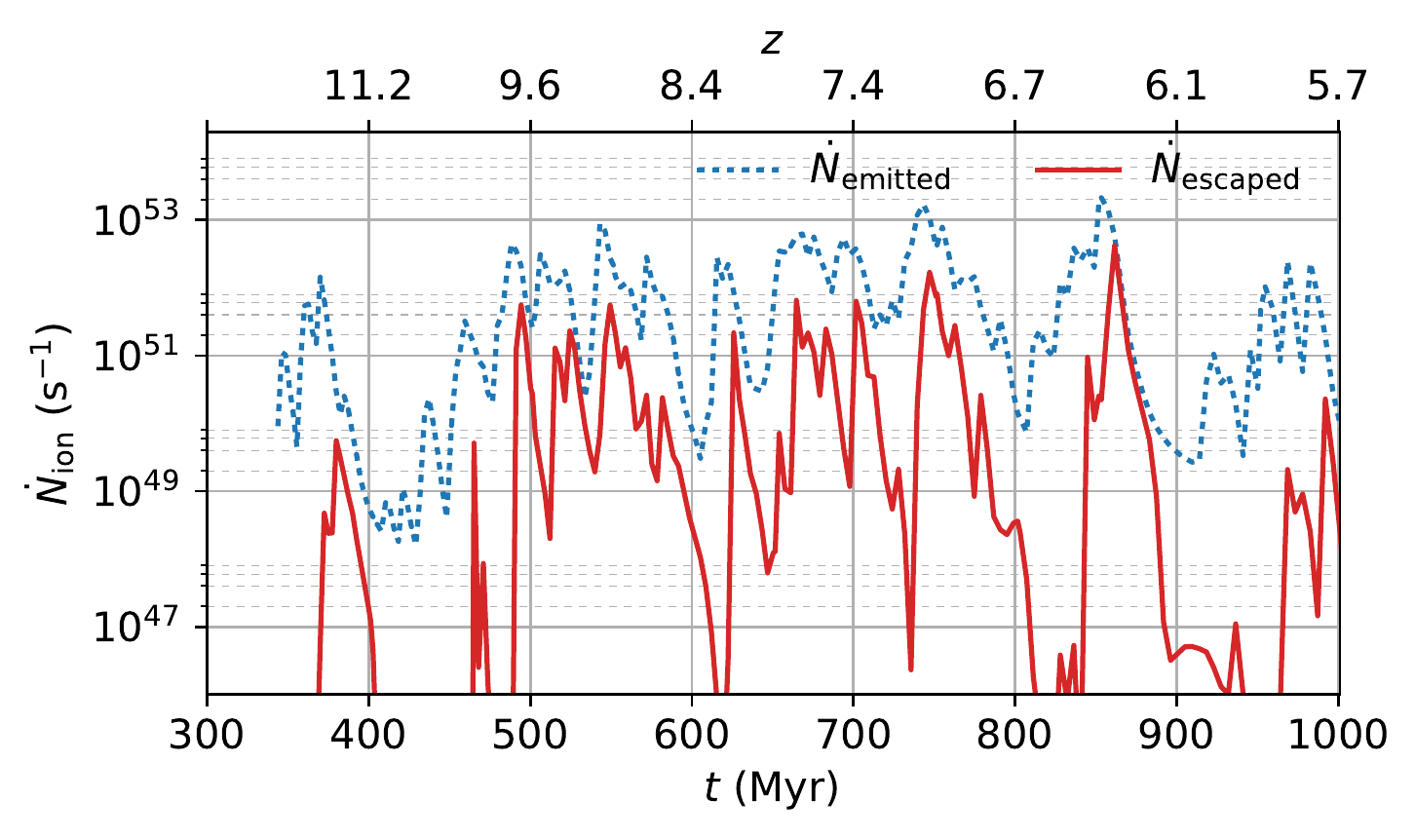}
  \caption{Photons emitted (blue) and escaping into the IGM (red) per second from the most massive halo. $\dot{N}_{\rm escaped}$ broadly follows $\dot{N}_{\rm emitted}$, but can drop significantly lower. Not all peaks in $\dot{N}_{\rm emitted}$ have corresponding peaks in $\dot{N}_{\rm escaped}$, meaning that not all episodes of star formation contribute to reionization.}
  \label{fig:nion}
\end{figure}

Fig.~\ref{fig:nion} presents the instantaneous, intrinsic ionizing luminosity of the most massive galaxy ($\dot{N}_{\rm emitted}$) in red, and the remaining luminosity that escapes the halo ($\dot{N}_{\rm escaped}$) in blue. The general trend is the same as for the evolution of the escape fraction: both the emitted and the escaped luminosity vary quickly over several orders of magnitude. However, we note that the phases during which $\dot{N}_{\rm escaped} \sim \dot{N}_{\rm emitted}$ (high escape fraction) do not necessarily correspond to peaks of the emissivity of the galaxy. For instance, right before $t \simeq 1$ Gyr, \fesc reaches 15\% and the photon production rate is roughly $\dot{N}_{\rm emitted} \simeq 10^{50}$ ionizing photons per second. The galaxy produces a similar amounts of photons around 920 Myr, but there \fesc barely reaches 0.1\%.
The maximum injection of ionizing photons into the IGM does not necessarily correspond to the peaks of star formation (and thus of intrinsic luminosity). Therefore, a galaxy experiencing a strong star formation episode will not always contribute significantly  to the ionizing budget of reionization.

Overall, if we assume that our simulated sample is reasonably representative of the low mass, high-$z$ galaxy population, we can expect that the fast paced variations of \fesc during the galaxies history result in a large scatter for the escape fraction for a homogeneous population of galaxies with e.g. the same mass. Indeed, the exact value of \fesc is highly dependent on the phase (pre-starburst, starburst or post-starburst) the galaxy is going through. Even at fixed \Mstar, a galaxy population will likely be out of phase, leading to a very different \fesc from galaxy to galaxy.

\subsection{Escape fraction estimator}
\label{sec:fescestimate}

\begin{figure}
  \centering
  \includegraphics[width=\linewidth]{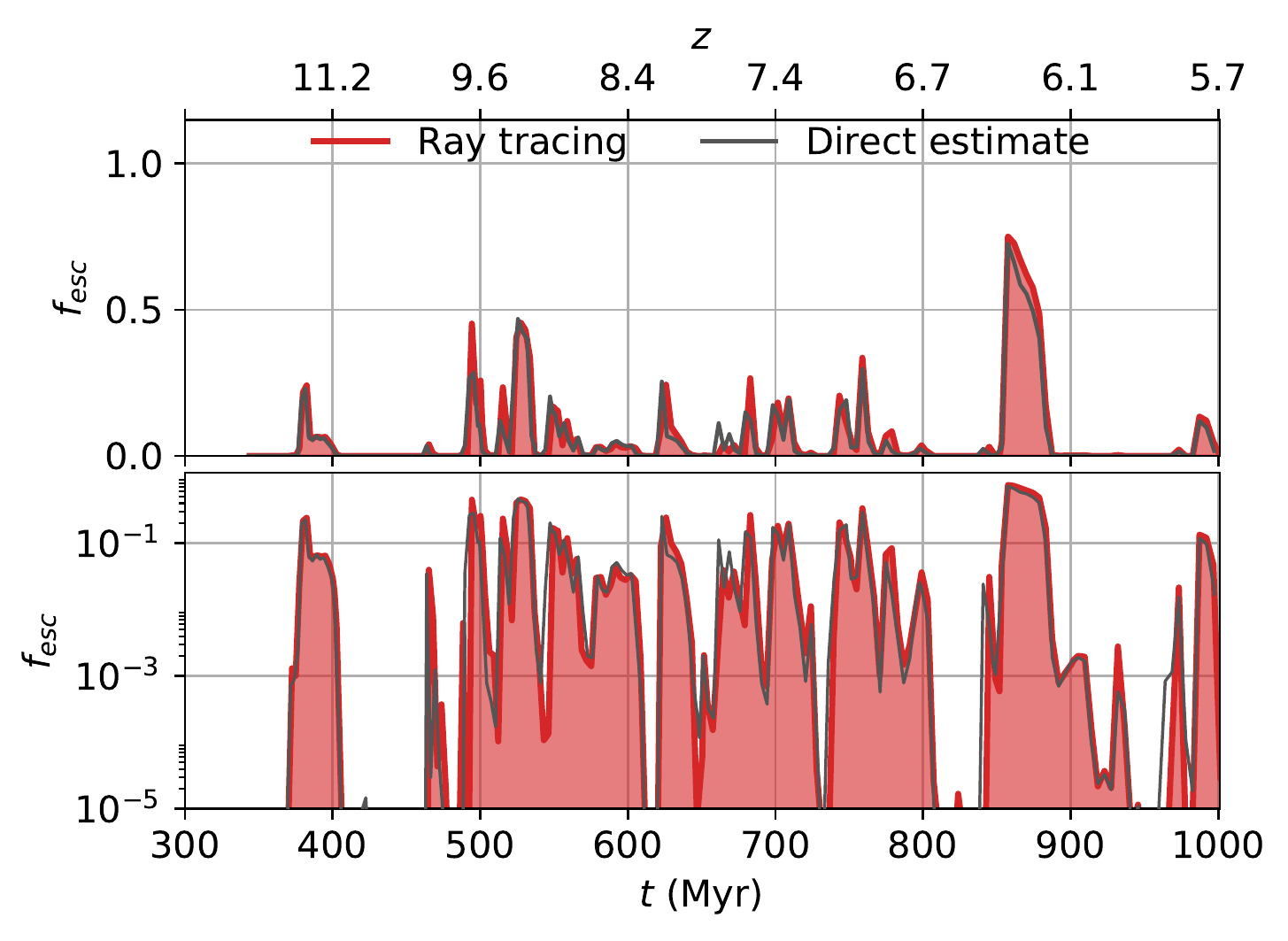}
  \caption{Comparison of the two estimators of \fesc: based on ray casting in red, and based on the local ionizing flux in black. The two estimators agree very well.}
  \label{fig:fescray}
\end{figure}

We introduced in Sect.~\ref{sec:fescmethod} a second estimator for the escape fraction using ray-tracing in order to test the robustness of our measurements of the escape fraction. 
We present in Fig.~\ref{fig:fescray} the angle averaged, luminosity-weighted escape fraction computed using eq.~\ref{eq:fescray} for the most massive halo. The red filled area represents the ray-tracing estimator of \fesc, while the solid black line is the flux estimator used in the previous figures. We plot \fesc both in linear (upper panel) and logarithmic scale (lower panel) to better illustrate the amplitude of the variations.

We could in principle expect some differences : the flux-based estimator assumes that all stars are at the centre of the halo, and may also give imperfect estimates as the moments method we use for the radiative hydrodynamics is known for being too diffusive \citep{Iliev2006, Iliev2009, Rosdahl2013}, leading to spurious leakage of radiation around opaque obstacles that should in principle cast a shadow. As we can see, this is not the case, and it is very reassuring that the two estimators are in very good agreement: the variations are seen at the same times and have very similar amplitudes.

We explain this by the fact that the stars are concentrated at the centre of the halo. Seen from the virial radius, the behaviour of the galaxy is very close to that of a central point source. The effect of the ray-crossing issue of M1 methods (i.e. when two opposed collimated beams would collide instead of cross each others) seems to be marginal, again because of the light sources are numerous and all very central.

\subsection{Directionality}
\label{sec:directionality}

\begin{figure}
  \centering
  \includegraphics[width=\linewidth]{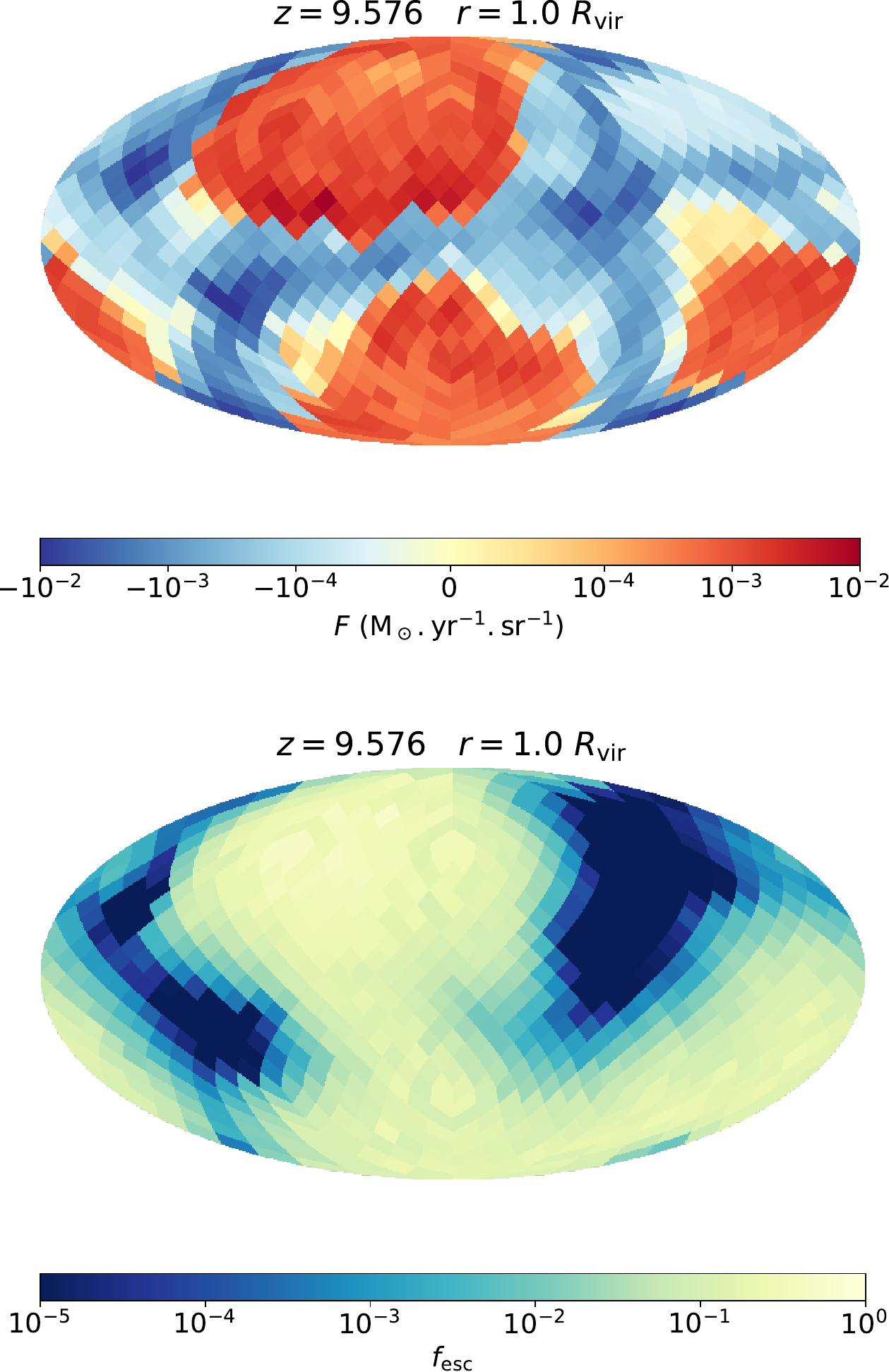}
  \caption{Full-sky Mollweide projection of the gas flows (upper panel) and of the escape fraction (lower panel) at $R_{\rm vir}$ from the centre of the most massive halo at $z \sim 9.6$, when $\fesc \sim 20\%$. In the upper panel, the gas mass flux $F$ is positive for outflowing gas and negative for infall.
    Photons mostly escape through channels carved by SN feedback.}
  \label{fig:direction}
\end{figure}

In the upper panel of Fig.~\ref{fig:direction}, we show the angular distribution of gas flows reaching the virial radius for a snapshot at $z \sim 9.6$, when $\fesc \sim 20\%$.
The wind seems to develop in mostly three directions (because of the Mollweide projection, the rightmost and leftmost patches look like they are disconnected), with a large part of the sky blocked by infalling gas. On the lower panel, the angular distribution of \fesc is displayed using the same projection. It shows clearly that radiation escapes the halo through channels created by the strong winds: the high-\fesc patches follow the same morphology as the outflows.

The importance of those high-outflow channels has been already studied e.g. by \citet{Fujita2003}, who found that the formation of outflows is a necessary condition for the creation of low column density direction through which ionizing radiation can escape the galaxy. This is in line with the more recent findings of \citet{Gnedin2008}, \citet{Wise2009}, \citet{Kim2013} and \citet{Paardekooper2015} who showed that the ionizing radiation escapes anisotropically, favouring low column density regions resulting from such outflows.

\subsection{Contribution to the ionizing budget}
\label{sec:budget}

As we have discussed in Sect.~\ref{sec:fesc_t}, at a given stellar mass, the escape fraction varies a lot and is not in phase with the evolution of the star formation rate. This results in situations in which a galaxy can have a high \fesc and still contribute only very little to the ionization of the IGM. In order to make a step forward in quantifying the actual contribution of small galaxies to the ionizing budget of the Universe, we compute the ionizing duty cycle of our galaxies. We define the duty cycle as the fraction of the time spent by all galaxies in a mass bin with $\dot{N}_{\rm escaped} \geq 10^{50}\ \mathrm{s}^{-1}$.
\begin{equation}
  \label{eq:duty}
  t_{\dot{N}_{50}} (\Mstar^i) = \frac{\sum_{\mathrm{galaxy}}\Delta t\left(M = \Mstar^i \middle| \dot{N}_{\rm escaped} \geq 10^{50}\ \mathrm{s}^{-1}\right)}{\sum_{\mathrm{galaxy}} \Delta t \left(M = \Mstar^i\right)},
\end{equation}
where $\Delta t \left(M = \Mstar^i\right)$ is the time spent by the galaxy in the mass bin $\Mstar^i$, and $\Delta t \left(M = \Mstar^i \middle| \dot{N}_{\rm escaped} \geq 10^{50}\ \mathrm{s}^{-1}\right)$ is the time spent in the mass bin $\Mstar^i$ with $\dot{N}_{\rm escaped} \geq 10^{50}\ \mathrm{s}^{-1}$. We define similarly $t_{\dot{N}_{48}}$.

\begin{figure}
  \centering
  \includegraphics[width=\linewidth]{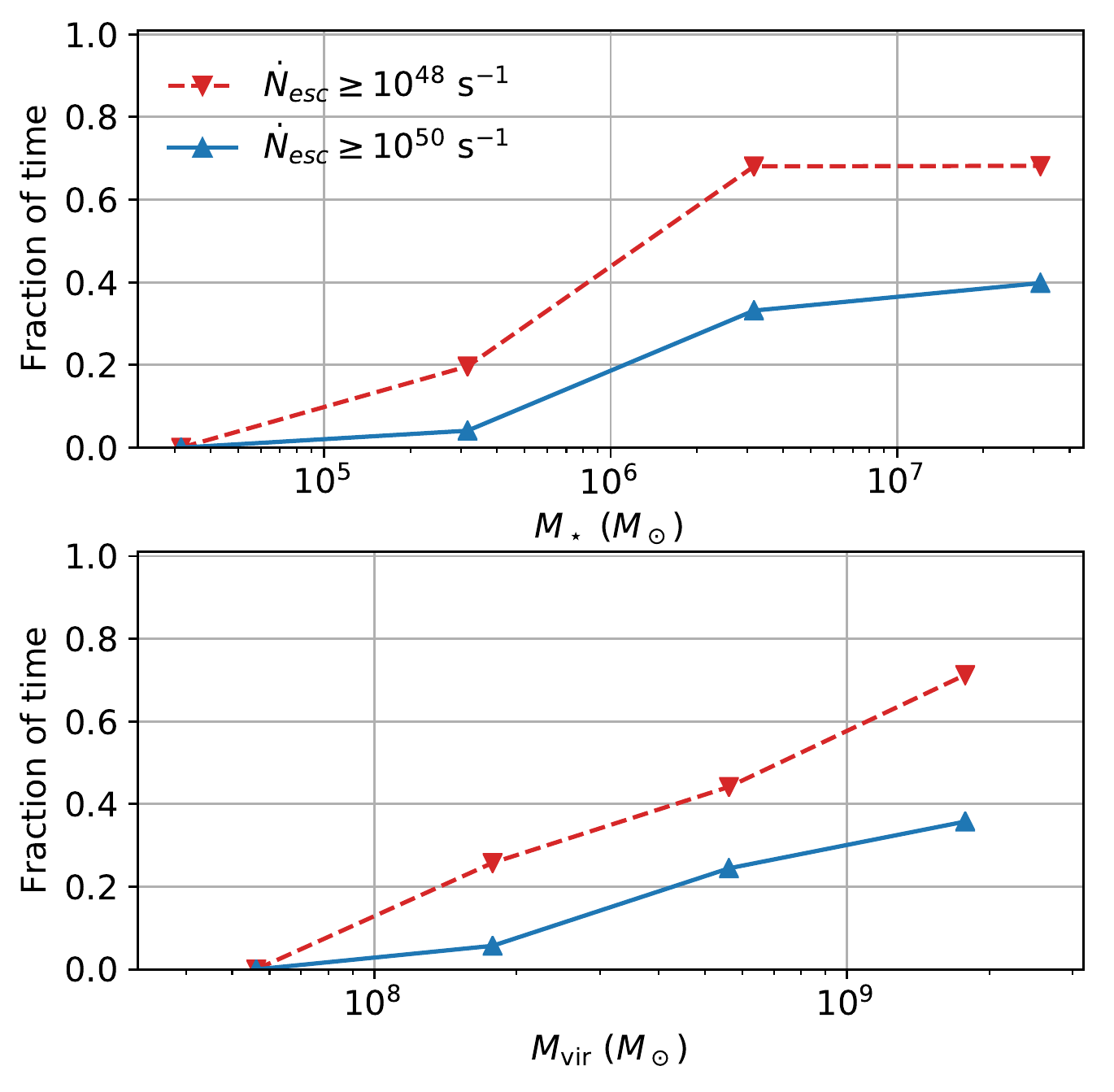}
  \caption{Ionizing duty cycle of the two most massive haloes, as a function of the galaxy (upper panel) and halo (lower panel) mass. The blue (red) dashed line shows the fraction of the time a galaxy of a given mass spends with $\dot{N}_{\rm escaped}$ higher than $10^{50}$ ($10^{48}$) photons per second. On average, a more massive galaxy spends more time leaking Lyman continuum photons.}
  \label{fig:dutycycle}
\end{figure}

We show these quantities in the upper panel of Fig.~\ref{fig:dutycycle}, for four regularly log-spaced mass bins, centred on $\log (\Mstar/\Msun) = 4.5, 5.5, 6.5$ and $7.5$, with $t_{\dot{N}_{48}}$ in red, $t_{\dot{N}_{50}}$ in blue. There is a clear trend that more massive galaxies tend to spend more time in a Lyman-leaking phase. At the very low-mass end of our plot, the duty cycle drops to zero: this corresponds to the first stellar population, formed in a single short episode, lasting around 20 Myr (e.g. see Fig.~\ref{fig:sfr_outflow}). When the first massive, Lyman-continuum producing stars end their lives, the production of ionizing photons drop rapidly, and is therefore quite low when the supernovae start to create clear channels. At these early times, the amount of photons reaching the IGM before the supernova phase is low as well. 
%%%
Similarly, the lower panel of Fig.~\ref{fig:dutycycle} shows the ionizing duty cycle as a function of halo mass, $t_{\dot{N}_{48}} (\Mvir)$ and $t_{\dot{N}_{50}} (\Mvir)$, for four regularly log-spaced mass bins centred on $\log (\Mstar/\Msun) = 7.75, 8.25, 8.75$ and $9.25$.
%%%
This is in strong disagreement with the recent results of \citet{Kimm2017}, who simulated the formation of mini-haloes and found that \fesc can be very high even before the first supernova. 
However, their study focuses on much lower mass haloes, reaching $\Mvir \leq 10^8\, \Msun$ at most. They also include a description of pop III stars and molecular hydrogen, which can lead to significant differences for haloes below the atomic cooling limit of $\Mvir \sim 10^8\, \Msun$.

On the other side of the mass spectrum, when $\Mstar > 10^6\Msun$, the duty cycle saturates, with $t_{\dot{N}_{50}} \simeq 40\%$ and $t_{\dot{N}_{48}} \simeq 70\%$. This happens when the galaxy starts to reach a more stable regime, with feedback-regulated star formation, but where the star formation is rarely completely shut off. This means that the gas in the galaxy will usually not have the time to settle down in between two starburst events, leading on average to a more steady production of ionizing photons.

\section{Local escape of ionizing radiation}
\label{sec:escape}

\begin{figure*}
  \centering
  \includegraphics[width=\linewidth]{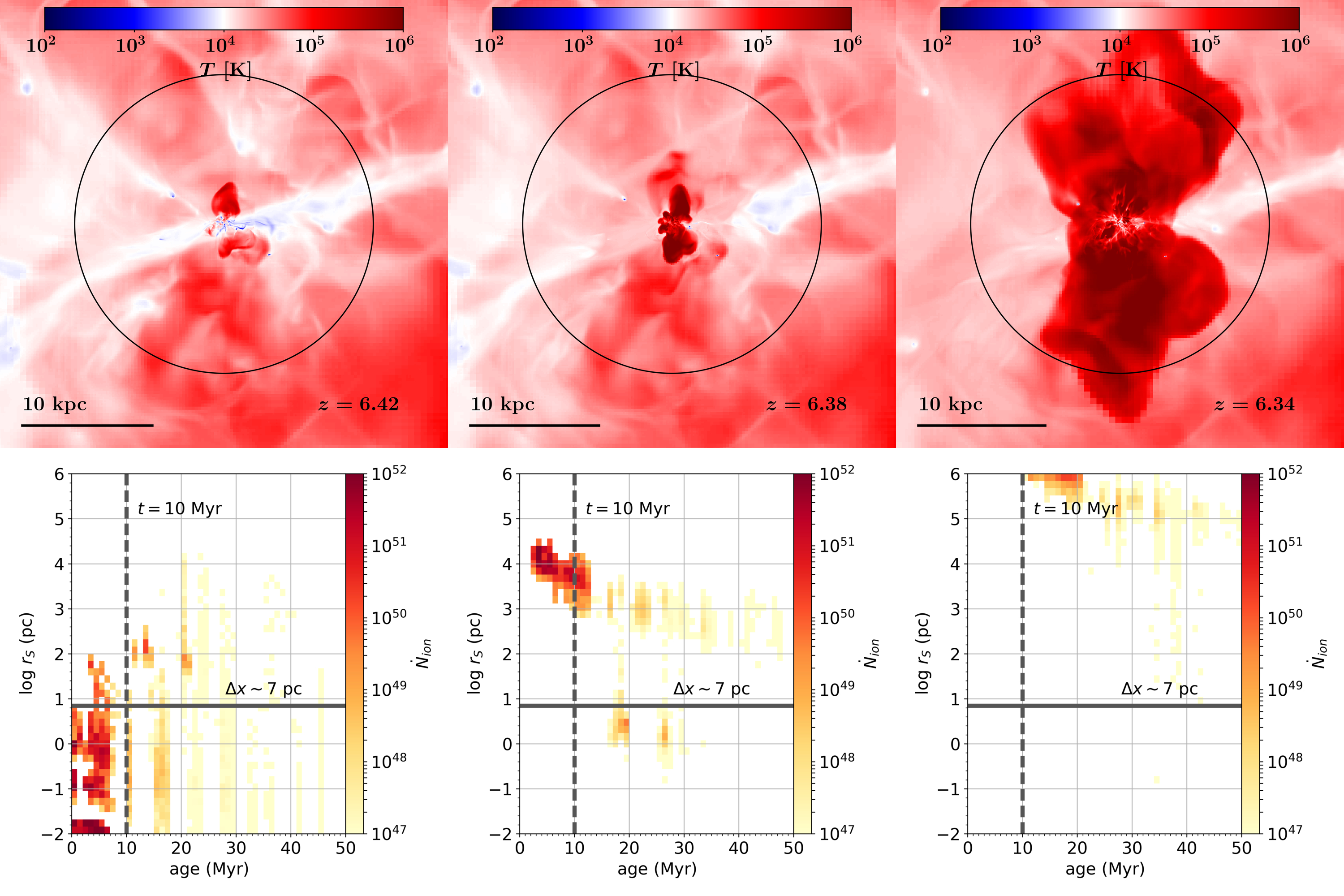}
  \caption{{\em Top row}: Temperature maps of the largest galaxy of our sample before (left), at the beginning (middle) and a few million years after (right) a massive feedback event. {\em Bottom row}: Luminosity-weighted distribution of the Str{\"o}mgren radius of each stellar particle (see text), as a function of the age of the particle. The thick horizontal black line indicates the size of the most refined cell. Areas below the line correspond to radiation absorbed inside the emission cell. Each pixel is colour-coded by the total ionizing luminosity.}
  \label{fig:environment}
\end{figure*}

In Sect.~\ref{sec:fesc}, we have found that the escape fraction is mainly regulated by the interplay between supernovae, gas accretion, and clustered star formation, and we have limited our analysis to the total escape fraction of each galaxy, computed at the virial radius. Because it is strongly linked with ISM-scale phenomena, we will now concentrate on the intra-halo processes leading to the escape of ionizing radiation. For this purpose, we will mainly focus on the most massive galaxy in our sample.

\subsection{Does radiation escape from the emission clouds?}
\label{sec:absorption}

Before reaching the IGM, the ionizing photons must travel through the complex distribution of gas inside and around the galaxy. \citet{Kimm2014, Ma2015, Paardekooper2015} showed that when \fesc is low, most of the photons are absorbed very locally, within 100~pc of their emission point, and that it is indeed crucial to resolve properly the ISM in order to study properly the escape of ionizing radiation. We find very similar results with our simulations, and we suggest that this is the reason behind the strong coupling between the supernova feedback and evolution of \fesc.

While the idea that clear channels for ionizing radiation from young stars is created by ionizing radiation itself is appealing, \citet{Geen2015} showed that depending on the structure of the cloud and the strength of the source, the \hii region may or may not expand beyond the boundaries of the cloud. We note that even with our high resolution, we cannot properly resolve the internal structure of these \hii regions. Nevertheless, we can still make a first step in the analysis of the local escape of photons from the star forming clumps. If we define the Str{\"o}mgren radius\footnote{\label{fn:stromgren}While the Str{\"o}mgren analysis only holds for an homogeneous medium with a static source, and assumes no backreaction of the radiation on the gas, we take it as a lower limit for the radius of the ionization front. Indeed, if the front expands further, it will do so quicker than the lifetime of the massive stars. Since we do not resolve properly the internal structure of star forming clouds, this is a first order approximation that allows analysis without introducing more free parameters.}
$r_S$ of an ionizing source as the radius of the sphere within which the rate of recombination is balanced by the ionizing luminosity, we can express $r_S$ as
\begin{equation}
  \label{eq:stromgren}
  r_S = \left(\frac{3}{4\pi}\frac{\dot{N}_{\rm ion}}{n_0^2\alpha_{\rm B}(T)}\right)^{\frac{1}{3}},
\end{equation}
where $\dot{N}_{\rm ion}$ is the rate of ionizing photons emitted by the source, $n_0$ is the density, and $\alpha_{\rm B}(T)$ is the case B recombination rate, given by \citet{Hui1997}. A young star particle (before 3-4 Myr) of 135 \Msun (our mass resolution) will yield approximately $4 \times 10^{48}$ ionizing photons per second. For a typical diffuse ISM density of $n_0 = 1$ cm$^{-3}$, this gives $r_S \sim 50$ pc, much larger than our typical cell size of 7 pc. However, inside a star forming cloud, the density can reach $n_0 = 1000$ cm$^{-3}$ in our simulations, resulting in $r_S \sim 0.5$ pc, meaning that the radiation should all be absorbed inside the emission cell.

While this depends on the inner structure of the cloud, this should in principle prevent the apparition of high escape fraction episodes for most star forming clouds. We argue that the leakage of ionizing radiation from such dense clouds is the result of supernovae in the neighbourhood of the young star cluster. The explosion removes gas from the cluster, effectively lowering the local gas density and thus making it possible for radiation to escape before the end of the few Myr lifetime of the massive stars.
In Fig.~\ref{fig:environment}, we compare three consecutive snapshots: right before a massive feedback event, right after, and when the outflow reaches the virial radius. The upper panel shows the temperature maps for the three snapshots, where the outflow can be traced by the expansion of the hot (dark red) region. In the lower panel, we present the corresponding luminosity-weighted distribution of Str{\"o}mgren radii for the star particles as a function of their age. Note that this assumes that each star particle is alone in its cell. Because stars form in clump, this will underestimate the effective Str{\"o}mgren radius of each star. Before the onset of supernovae, we see that most of the ionizing radiation is produced by young stars embedded in dense cells (the red region in the bottom left corner of the plot), resulting in $r_S$ much smaller than the cell size. In the middle column, right after the onset of supernovae, the ionizing emissivity is dominated by young stars for which $r_S$ is several thousands time larger than the cell size (the dark red spot at $\sim$ 5 Myr and $\log r_S \sim 4$). While the value of $r_S$ is approximate (e.g. because the ISM is not homogeneous), this means that a lot of radiation will escape from the galaxy. The rightmost column shows the expansion of the outflow: the stellar population is ageing, and while radiation continues to leak (meaning a high \fesc), the net ionizing emissivity is much lower. We can match that sequence to Fig.~\ref{fig:fesc_sfr_out} and \ref{fig:nion}: at $\sim$ 850 Myr, there is noticeable burst of star formation, followed by the development of a large outflow. At the beginning, the intrinsic ionizing emissivity is high, but \fesc is very low. Then, just as the outflow rate starts to increase, the escaped emissivity rises, and $\fesc \sim 60\%$. In the final stages of the succession of events, the star formation rate drops and there are no more young massive stars left. Both the intrinsic and escaped emissivity drop, so \fesc remains high, of the order of 50\%.

\begin{figure}
  \centering
  \includegraphics[width=\linewidth]{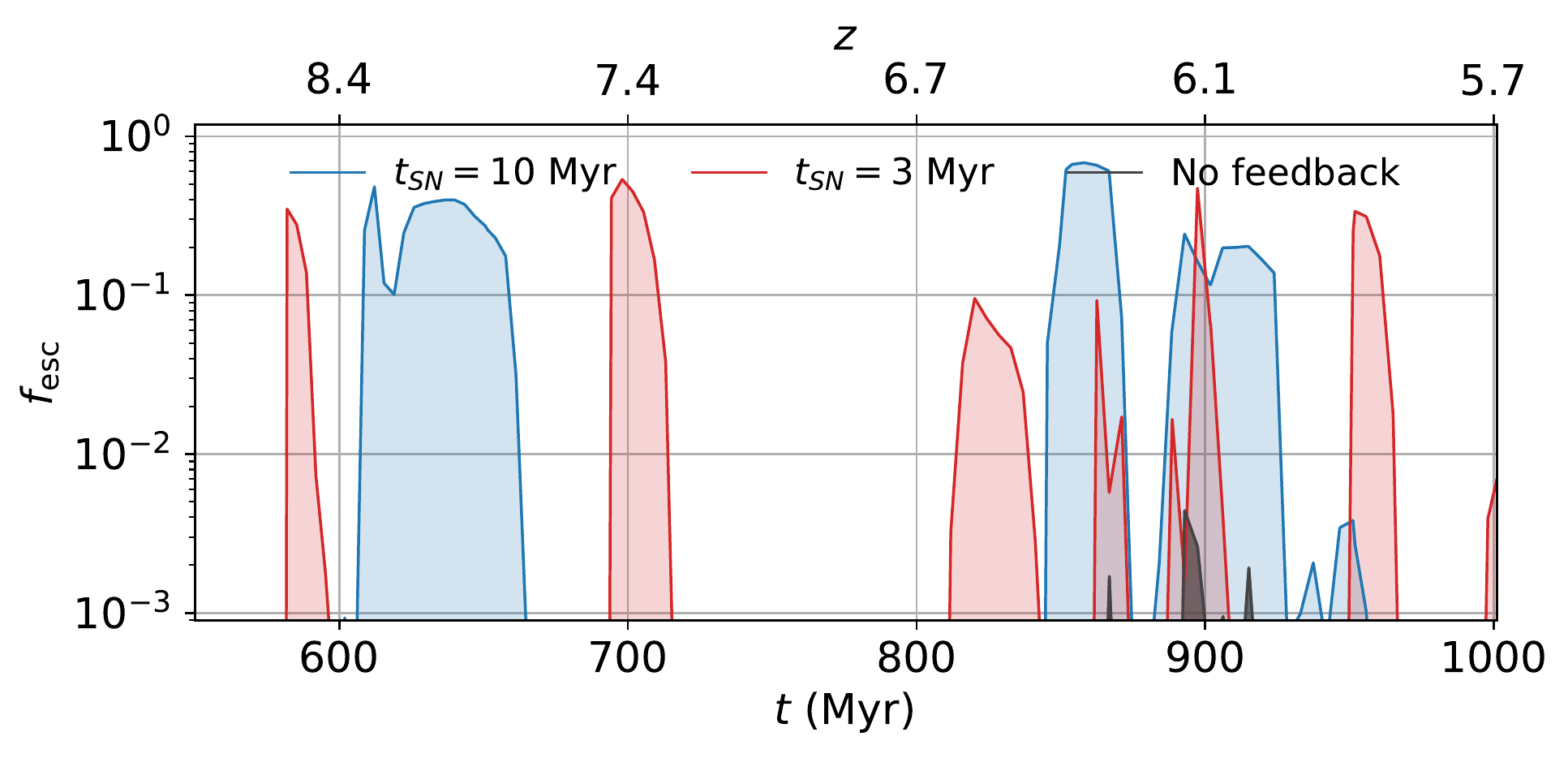}
  \caption{Evolution of \fesc for the medium halo with SN feedback and a time delay $t_{SN}$ of 10 Myr (3 Myr) between star formation and the supernova in blue (red), and without feedback in black.}
  \label{fig:comparefb}
\end{figure}

As an additional line of evidence that SN feedback is a crucial element of the journey of ionizing photons from the stars to the IGM, we present in Fig.~\ref{fig:comparefb} a comparison of \fesc for three simulations of the intermediate mass halo with variations on the feedback. The blue curve shows the fiducial simulation presented before, the red curve corresponds to a run where the delay $t_{SN}$ between star formation and the supernova phase has been reduced to 3 Myr, and the black curve corresponds to a case without supernovae.
Interestingly, the total mass of stars formed in the simulations with feedback is very close, with less than 15\% difference after 1 Gyr.
For the two simulations with feedback, the evolution of \fesc is qualitatively similar: rapidly alternating, and reaching values as high as 30\% to 50\%. Even if it is difficult to quantitatively assess the influence of the delay because of the randomness of the starburst events, these two runs present behaviours strongly contrasting with the no-feedback run, for which almost no radiation escape. In that last case, even though much more stars are formed over the course of the simulation, \fesc is always below 1\%. This indicates that, at least at the $\simeq 10$ pc resolution of our simulations, radiation feedback from the young stars in not strong enough to ionize the star-forming cloud, and thus that it is SN feedback that permits radiation to escape.

\begin{figure}
  \centering
  \includegraphics[width=\linewidth]{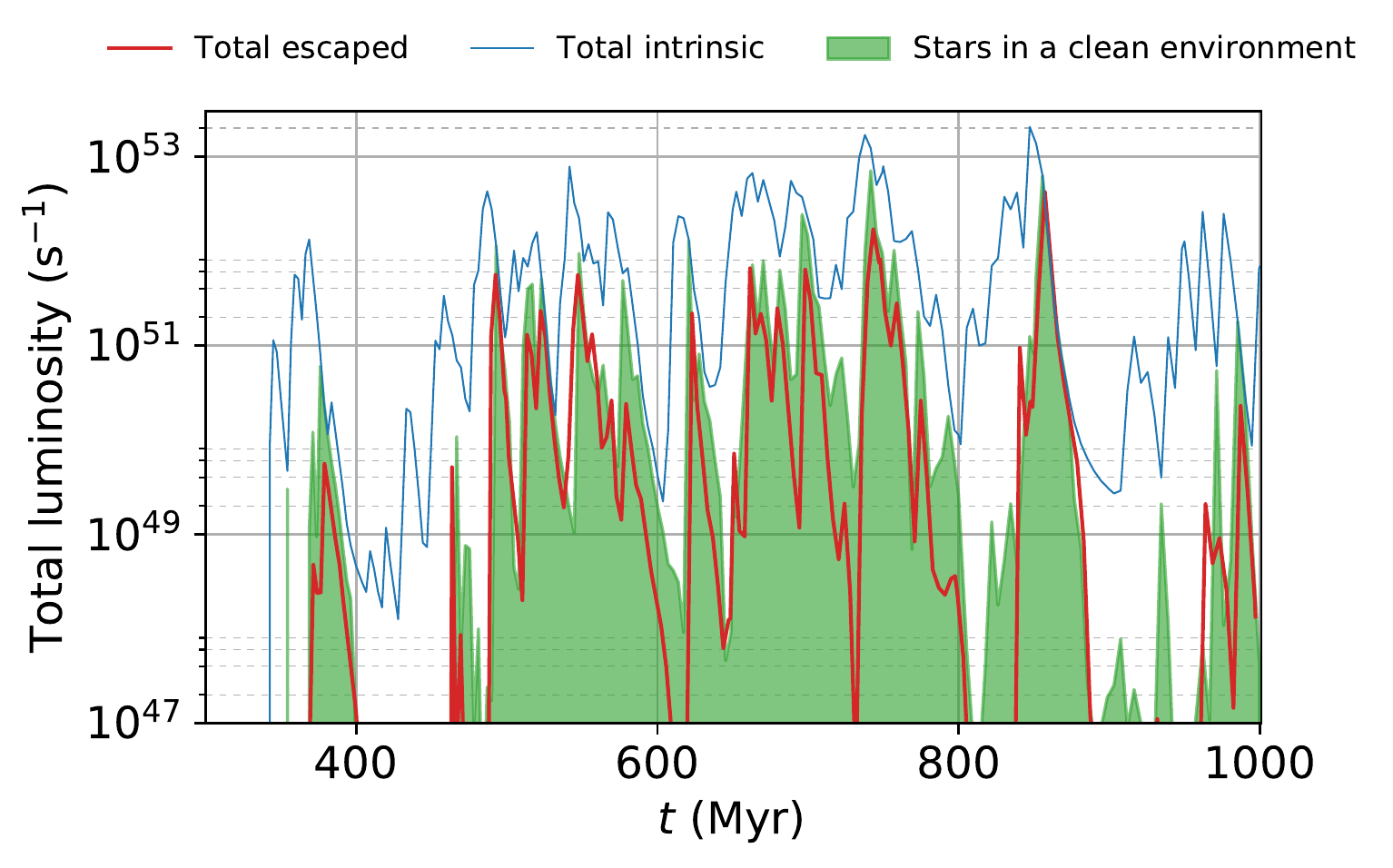}
  \caption{Luminosity contributed by stars in a clear environment (filled green) compared to the total emitted luminosity (blue). The escaped luminosity is indicated by a solid black line and is closer to green curve than to the total emitted luminosity.}
  \label{fig:clearfraction}
\end{figure}

This analysis indicates that the escape of ionizing radiation is mostly a local phenomenon, controlled by whether or not the photons can escape the cloud in which they have been emitted. Since the typical size for molecular clouds is of the order of a few tens of parsecs, we can consider a star particle to be in a clear environment if $r_S \geq 5 \Delta x \simeq 35$ pc. In Fig.~\ref{fig:clearfraction}, we compare the total intrinsic ionizing luminosity (in blue) to that of stars in such a clear environment (in filled green). We show the escaped luminosity as a red line. Overall, high escape fraction episodes happen in phases where the ionizing luminosity is dominated by stars in a clear environment. We can note that in most cases, the evolution of the luminosity contributed by stars in a clear environment follows closely that of the escaped luminosity, strengthening our scenario in which the escape of ionizing radiation is mainly regulated on the cloud scale.

\begin{figure}
  \centering
  \includegraphics[width=\linewidth]{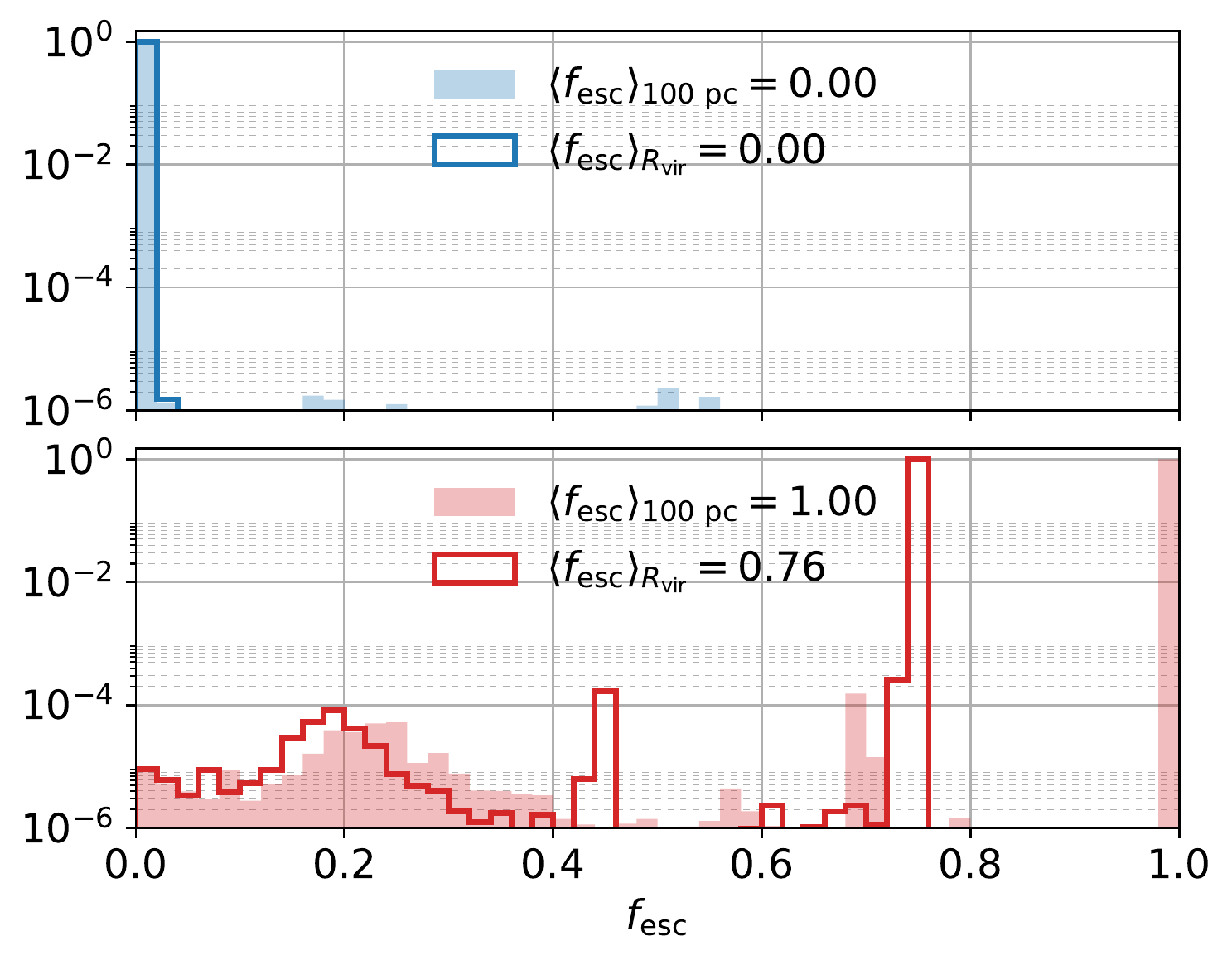}
  \caption{Distribution of the luminosity-weighted individual escape fraction from young stars for a snapshot where the global \fesc is low on the upper panel, and high on the lower panel. The thick lines denote the individual escape fraction at the virial radius, and the thin lines show the local escape fraction at 100 pc.}
  \label{fig:fesclocal}
\end{figure}

We continue this analysis in Fig.~\ref{fig:fesclocal}, where we compare the escape fraction on a local scale (filled areas) and at the virial radius (thick solid lines), for a snapshot in which the galaxy is leaking ionizing radiation (lower panel) and for another where no radiation escapes the halo (upper panel). For both panels, \fesc has been estimated using the ray-tracing method described in Sect.~\ref{sec:fescmethod}, and we plot the luminosity-weighted distribution of the transmissivity per star particle $i$ averaged over all directions $j$, defined as $\bar{T}_i = \langle e^{-\tau_{\hi}^{i,j}} \rangle_{j}$, where the optical depth is integrated over both 100 pc and $1\ R_{\rm vir}$. As depicted on the upper panel, it is clear that when no radiation escapes the halo, it is because all the photons are absorbed locally, within 100 pc of their emission site. When the cloud has been destroyed and all the photons escape locally (lower panel), most of them will be able to reach the IGM. This is very consistent with previous results of e.g. \citet{Kimm2014, Ma2015, Paardekooper2015}.

\subsection{Absorption within the halo and in the CGM}
\label{sec:CGM}

Simulations such as the ones we present in this work can be used to calibrate large scale models of reionization, or analytical estimates of the history of reionization. The key figure in these models is usually the angle averaged halo escape fraction, which represent the amount of ionizing photons that will escape the DM halo they originate from. This is typically the view that is adopted in semi-analytical models.
In the previous sections, we have argued that the escape of ionizing radiation is mostly regulated on a local scale by various stellar feedback processes.
A fraction, however, is still absorbed within the halo. This can be either because of intervening clouds or just because the halo is not fully transparent. For example, in the case of the snapshot corresponding to the the lower panel of Fig.~\ref{fig:fesclocal}, about $\sim 25\%$ of the hydrogen in the halo is neutral, which be responsible for the absorption in the halo.
The LyC leaking episodes typically correspond to the development of strong galactic winds, that significantly alter the gas distribution inside and around the galaxy, especially in low mass galaxies where SN feedback is thought to be the most relevant \citep[see e.g.][]{Teyssier2013}. These winds can reach several virial radii, somewhat challenging the implicit assumption that the halo is a black box which radiative efficiency can be described with only the \fesc parameter.
We attempted to use $\fesc(r)$ radial profiles to determine whether or not the virial radius is characteristic of the escape of radiation (e.g. does \fesc reach a plateau at the virial radius?), and it appears that $R_{\rm vir}$ is probably not a relevant distance for photon propagation. This is not surprising: we have shown that the escape of radiation is a local process, and gaseous haloes around galaxies extend beyond $R_{\rm vir}$ anyway.

\begin{figure}
  \centering
  \includegraphics[width=\linewidth]{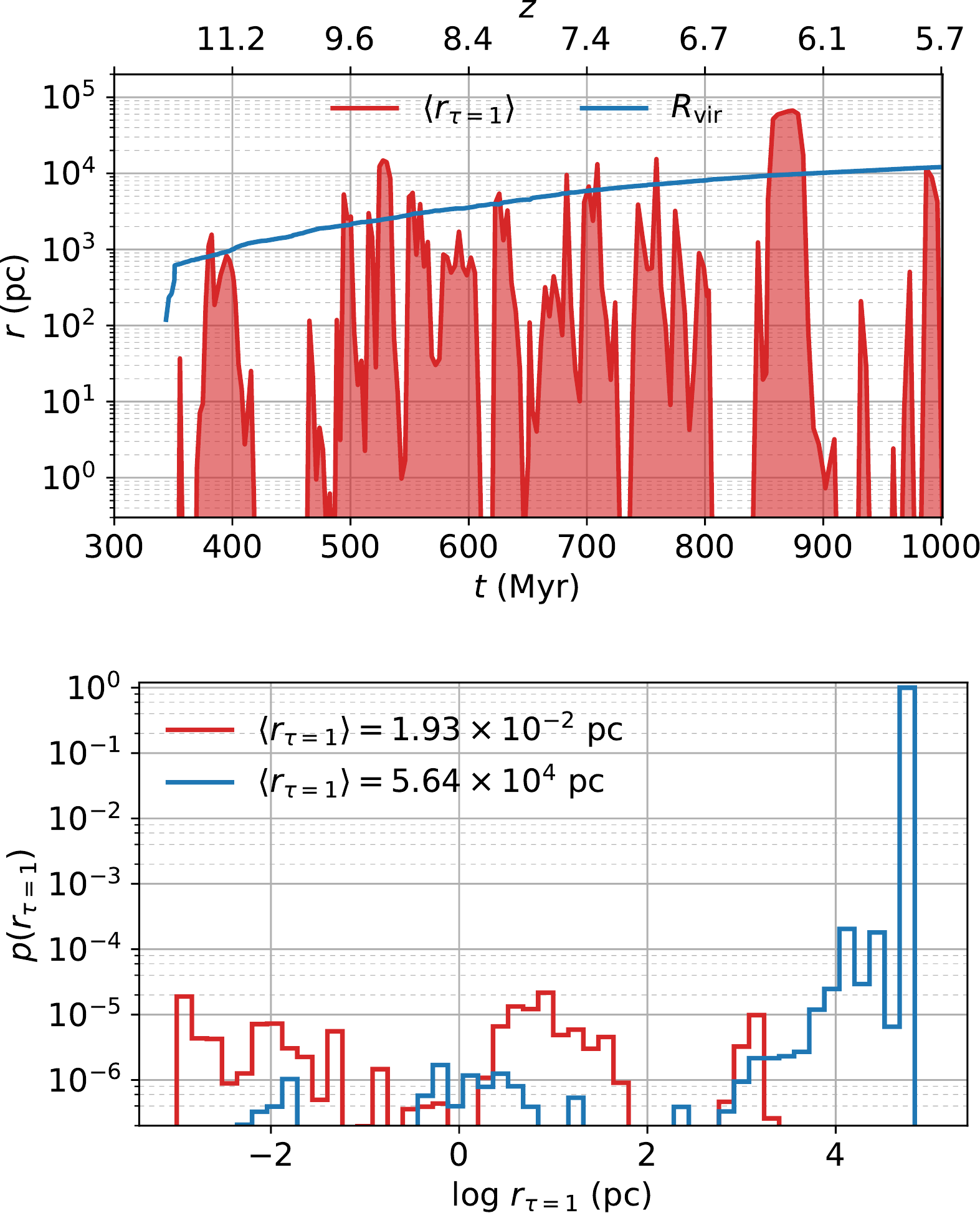}
  \caption{\textit{Upper panel}: Comparison of the typical absorption distance $r_{\tau=1}$ (in red) to $R_{\rm vir}$ (in blue). When $r_{\tau=1} > R_{\rm vir}$, most of the photons can escape the halo and reach the IGM. \textit{Lower panel}: Distribution of the angle-averaged $r_{\tau=1}$ around individual star particles for the two snapshots of Fig.~\ref{fig:fesclocal}.}
  \label{fig:r_combine}
\end{figure}

From the viewpoint of the ionizing sources, it makes more sense to elucidate at which distance photons are absorbed. With this in mind, we employ the same ray-tracing technique we described in Sect.~\ref{sec:fescmethod} with 192 rays per star particle, but instead of computing the optical depth $\tau$ at a given distance, we evaluate the distance $r_{\tau=1}$ at which the optical depth reaches $\tau = 1$. We compare in the upper panel of Fig.~\ref{fig:r_combine} the luminosity-weighted average $r_{\tau=1}$ (in red) to the virial radius of the halo (in blue). The lower panel presents a comparison of the distribution of this characteristic scale for the two snapshots already illustrated in Fig.~\ref{fig:fesclocal}, employing a higher resolution (we used 3072 rays per particle, corresponding to a \textsc{HEALPix} level of 4).
We note that the outputs where the average $r_{\tau=1}$ is high correspond to the LyC leaking episodes, and ultimately reaching the peak of \fesc when $r_{\tau=1} > R_{\rm vir}$. During these phases, there is very little absorption within the halo. The blue histogram corresponds to a snapshot at $t \simeq 850$ Myr, where \fesc reaches its highest value, around 75\%. The photons emitted by almost every star particle travel more than ten times farther than the virial radius before being absorbed. On the contrary, the red histogram corresponds to an episode of very low \fesc (at $t \simeq 815$ Myr), and on average, the radiation is absorbed within less than 0.01 pc, well inside the emission cell.
Looking again at the upper panel, we see that for most of the time, the typical absorption distance reaches 10 -- 1000 pc.
This corroborates our findings that channels created by SN explosions favour the escape of radiation. Unless the galaxy undergoes a massive, coordinated feedback event, there will still be absorption within the halo, even outside of the emission cloud.

\section{Observability of small galaxies}
\label{sec:observations}

\begin{figure}
  \centering
  \includegraphics[width=\linewidth]{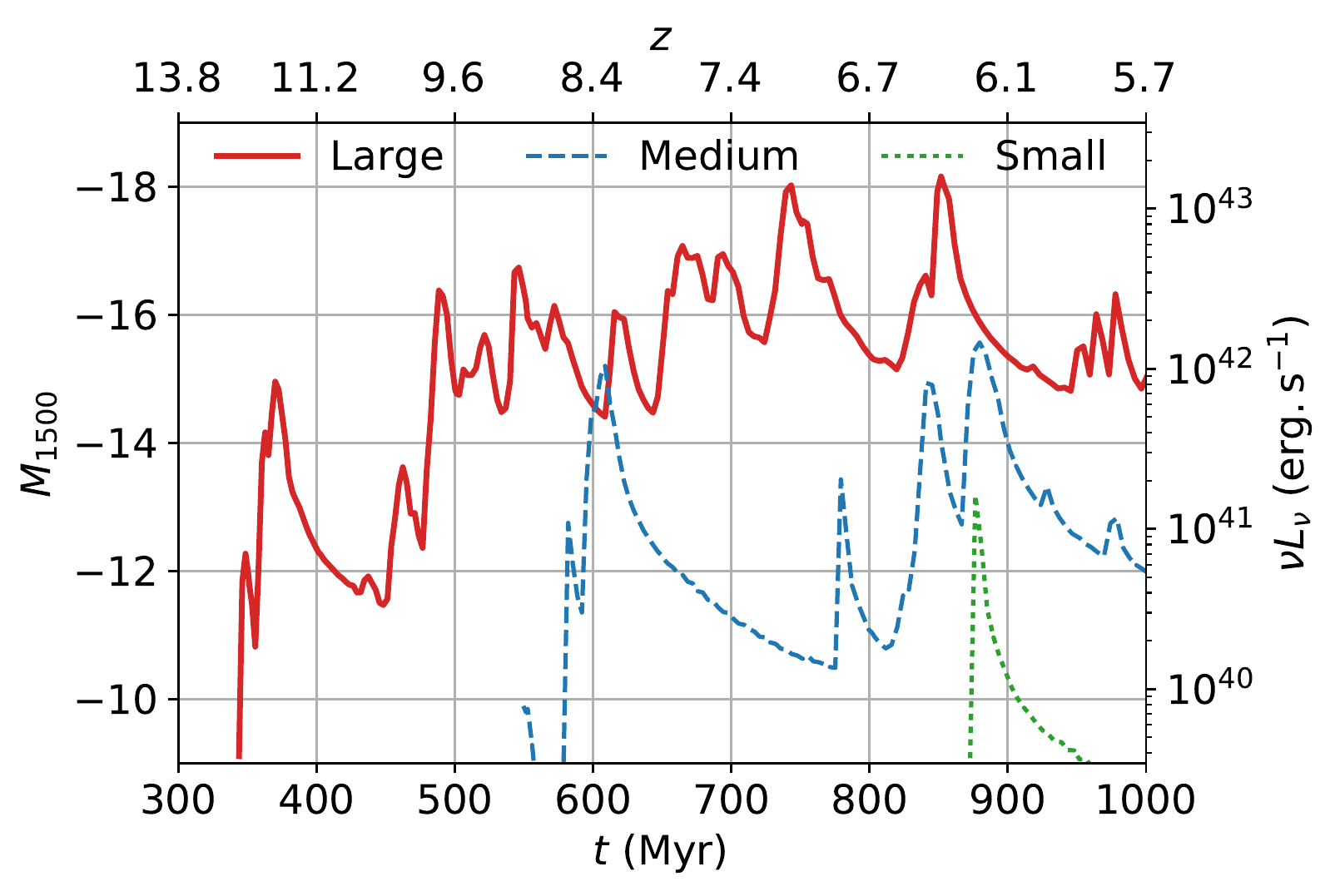}
  \caption{Evolution of the UV magnitude at $1500\, \angstrom$ for the three simulated galaxies. The variability follows that of the star formation rate, albeit on longer timescales.}
  \label{fig:MUV}
\end{figure}

Due to the opacity of the high redshift IGM to ionizing radiation, very deep surveys even with the next generation of telescopes such as the \emph{James Webb Space Telescope} (\emph{JWST}) will not be able to directly measure the ionizing flux coming from the sources of reionization. It is therefore necessary to assess the non-ionizing properties of such galaxies at frequencies that will be observed by these instruments.
For this purpose, we compute the non-ionizing, rest-frame UV magnitude around $1500\, \angstrom$ for the three galaxies of our sample. The luminosity of each star particle is given by the models of \citet{Bruzual2003} and is a function of its age and metallicity, but we do not include any dust absorption. To compute the luminosity of the galaxy, we sum the luminosities of each star particle within the virial radius. We present in Fig.~\ref{fig:MUV} the evolution of the absolute UV magnitude expressed in the AB magnitude system \citep{Oke1983}.
Overall, the peaks of UV luminosity broadly follow the episodes of star formation, with a much slower decline.
We find that the most massive galaxy can reach absolute UV magnitudes as high as $\muv \lesssim -18$, but spends most of its time at $\muv \gtrsim -15$. This lower limit is comparable to the deepest surveys using gravitational lensing to probe the very faint end of the luminosity function at high redshift \citep[e.g.][]{Atek2015}, and will be within the reach of the next generation of surveys using \emph{JWST} as proposed e.g. by \citet{Finkelstein2015}. Recently, \citet{Huang2016} found a Lyman-$\alpha$ emitting galaxy at $z \sim 7$ with a detectable UV continuum around $\muv \lesssim -18$ at $1600 \angstrom$. They inferred a mass of $\Mstar \sim 1.6 \times 10^7 \Msun$ and a star formation rate of $1.4 \Msun.\mbox{yr}^{-1}$. This is very similar to our most massive galaxy in its bursting phase, hinting that we are already starting to observe galaxies just like the ones presented in this work.

\begin{figure}
  \centering
  \includegraphics[width=\linewidth]{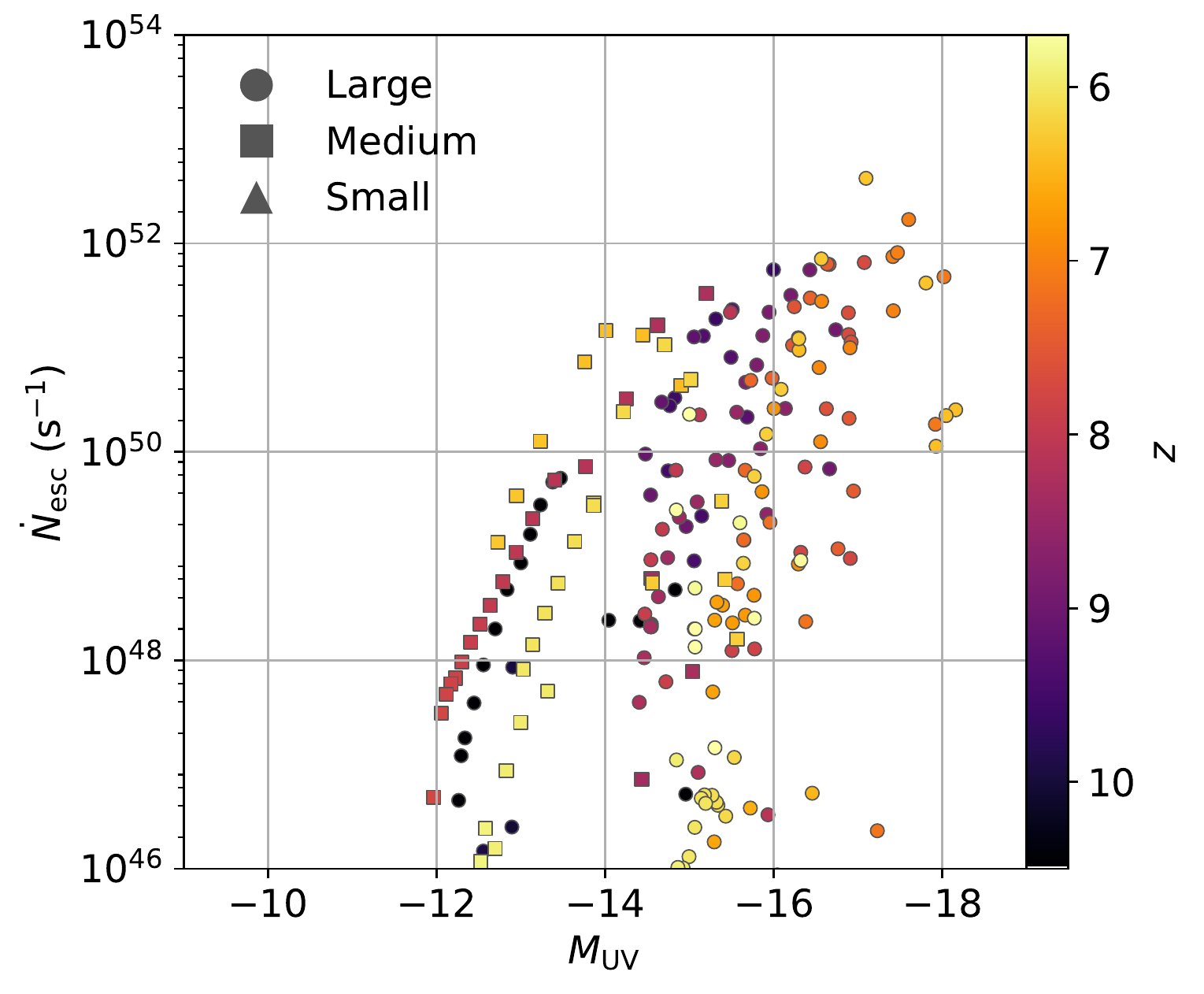}
  \caption{Comparison of the UV luminosity and the (escaped) ionizing luminosity for each snapshot of the three simulations. The redshift of the snapshot is indicated by the colour coding. The apparent lack of correlation is explained the much faster decrease of ionizing luminosity compared to non-ionizing UV.}
  \label{fig:MUV_nesc}
\end{figure}
However, Fig.~\ref{fig:MUV} shows that there is a large variability in the UV magnitude, and we have seen in Sect.~\ref{sec:fesc} that the escape of ionizing radiation is also highly variable following the star formation episodes. The correlation between the escaped ionizing luminosity $\dot{N}_{\rm esc}$ and the UV magnitude is relatively poor, as is illustrated in Fig.~\ref{fig:MUV_nesc}. The figure compares the UV magnitude and $\dot{N}_{\rm esc}$ for each snapshot of all three simulations, the colour coding indicates the redshift of the snapshot. We see that at any time when a galaxy shines at $\muv \sim -16$ in the UV, its ionizing luminosity spans over more than six orders of magnitude, ranging between $10^{46}$ and $10^{52}$ ionizing photons per second.
This lack of correlation can be explained by two factors: first, there is a delay between the peak of ionizing emissivity and escape of radiation due to the necessity to pierce the ISM, and second, the ionizing luminosity decreases much faster than the UV luminosity as a function of stellar age. This results in episodes with star formation (meaning high \muv and intrinsic ionizing emissivity) but very small \fesc, leading to a low $\dot{N}_{\rm esc}$.

\begin{figure}
  \centering
  \includegraphics[width=\linewidth]{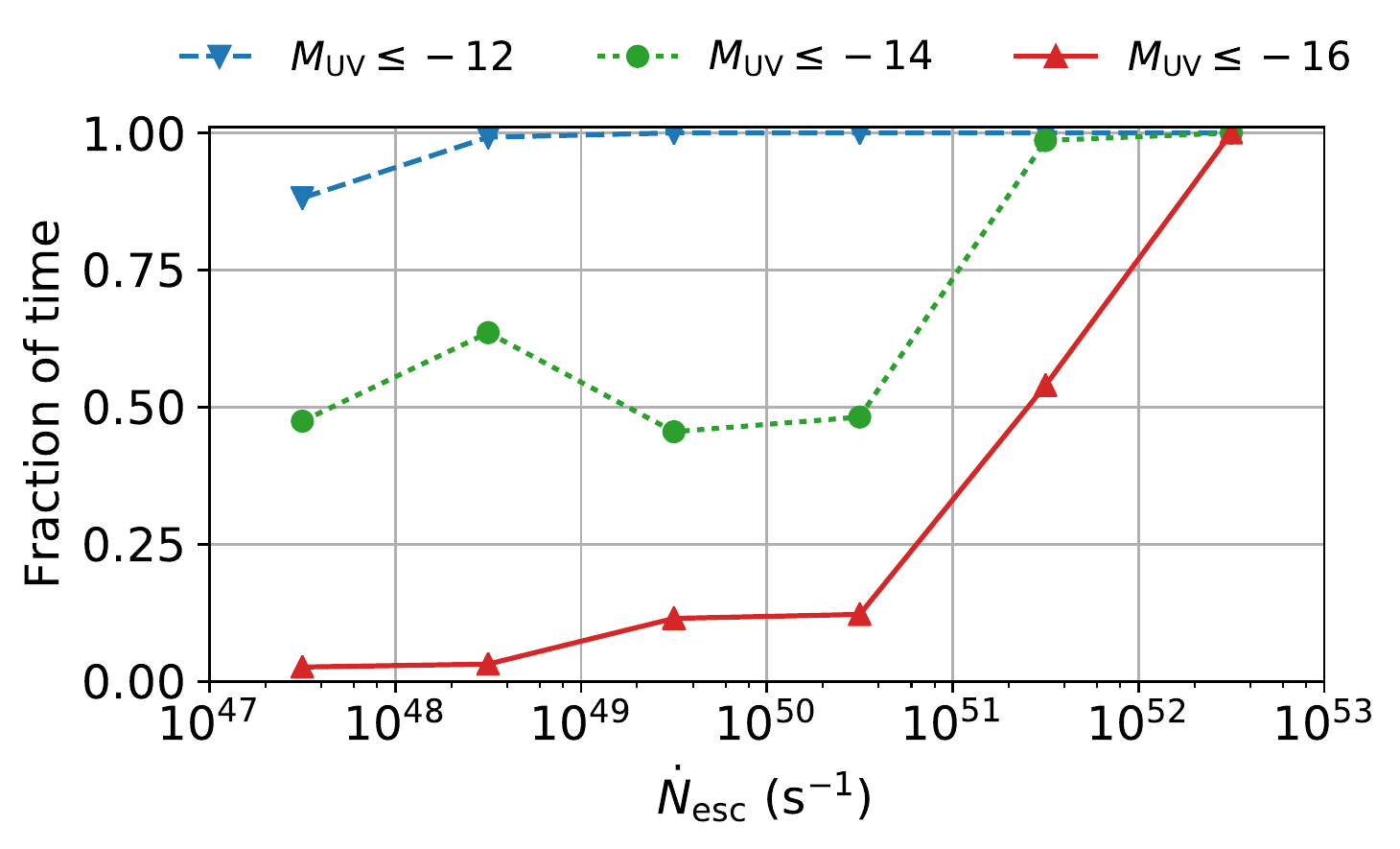}
  \caption{Estimation of the completeness of a sample limited by UV magnitude. The lines show the fraction of time that a galaxy emitting a given ionizing luminosity spends at a magnitude above -12, -14 or -16 (respectively in blue, green or red).}
  \label{fig:duty_MUV}
\end{figure}
Conversely, this has consequences for the selection of Lyman-leakers: a galaxy releasing at least $10^{50}$ ionizing photons per second in the IGM can be as bright as $\muv \simeq -18$ or fainter than $\muv \simeq -14$.
Fig.~\ref{fig:duty_MUV} shows this selection effect more quantitatively using a representation similar to Fig.~\ref{fig:dutycycle} : it shows the fraction of the time that a galaxy contributing $\dot{N}_{\rm esc}$ spends shining brighter than a magnitude of -12 (in blue), -14 (in green) or -16 (in red). Assuming that the full set of snapshots from our simulations represents correctly a population of faint galaxies, this can be read as some sort of completeness function. For instance, let us focus on galaxies emitting between $10^{49}$ and $10^{51}$ ionizing photons per second, which represent the bulk of the galaxies in the mass range considered in this work, according to \citet[][see their fig.~5]{Kimm2014}. A survey as deep as $\muv = -12$ would select all of them, while a survey limited at $\muv = -16$ would only see 10\% of them. This is of course a very rough estimate and should not be taken as a prediction for future surveys. Nevertheless, this resonates with the recent study of \citet{Hartley2016}, who used semi-analytical modelling and halo matching techniques to estimate the impact of the burstiness of the sources of reionization. They found that models with bursty star formation predict brighter galaxies at fixed halo mass.

We have so far only discussed quantities integrated in all directions, while we have shown in Fig.~\ref{fig:direction} that the escape of ionizing radiation can be very anisotropic. This will naturally lead to a larger scatter at fixed UV magnitude. The existence of preferred directions of escape will also result in anisotropic \hii regions, and this could affect the visibility of Lyman-$\alpha$ photons at high redshift. However, since the Lyman-$\alpha$ line is resonant, a proper treatment with dedicated radiative transfer is necessary. We plan to address this in a further study.

\section{Discussion}
\label{sec:discussion}

\subsection{On the importance of resolving the ISM}
\label{sec:discuss-ISM}

We have shown in this study that a large fraction of the photons emitted by young stars is absorbed in their immediate neighbourhood, corroborating the finding of \citet{Kimm2014} and \citet{Paardekooper2015}, who argued that the escape of radiation is mostly governed at the cloud scale. Similarly, \citet{Ma2015} found that the time-dependent, multiphase structure of the ISM in the vicinity of stars is an important element that regulates the escape of ionizing radiation. Earlier, \citet{Ciardi2002}, \citet{Clarke2002}, and then \citet{Fernandez2011} discussed how the porosity or clumpiness of the ISM affects \fesc. We wish now to discuss several aspects of the ISM scale physics, and how they would impact our results.

\subsubsection{Inner structure of molecular clouds}
\label{sec:inner-struct-molec}

Before going any further, we would like to discuss the question of the resolution of our series of simulation. Star formation being a process that occurs on scales well below the parsec, no cosmological simulation will be able to resolve the formation of individual stars in the near future, and even though the available computing power continues to increase, current simulations are limited. But what resolution is high enough? \citet{Wise2009} suggest that a resolution of 0.1 pc is needed to resolve the early phases of the expansion of the ionization front, but this is unattainable for the current generation of cosmological simulations. \citet{Rosdahl2015a} reached the rather depressing conclusion that, with a standard star formation model, \hii regions can never be properly resolved, just due to the fact that there is a limited amount of gas available in a cell for star formation. In light of this, we limit ourselves to a maximum resolution of $\Delta x \sim 7$ pc. This allows us to start to resolve the internal structure of our galaxies, which is mandatory to describe the sites where radiation gets absorbed.

Recently, \citet{Kimm2017} presented a study of the escape of radiation from mini-haloes, reaching the extremely high resolution of 0.7 pc. While their setup is similar to ours, they find that on some occasions, photons manage to escape even before the star particles reach their supernova phase. They interpret this as a sign that radiative feedback from the young stars can disrupt the star-forming clouds very early. We do not see this happening in our simulation, even though we include most of the relevant processes. Indeed, \citet{Kimm2017} find that the main channel of radiative feedback is through photoionization, which we also include in this work. This is in part due to our resolution, as we do not resolve the inner structure of the clouds. 
Most importantly, they focus on much lower mass haloes (reaching a maximum mass of $10^8\, \Msun$) for which they include the feedback from massive pop III stars. In this study, we only considered more massive haloes above the atomic cooling limit.

Using high resolution simulations of molecular clouds with radiative feedback, \citet{Dale2012} found that only a fraction of the photons emitted during the lifetime of a cloud will escape to the more diffuse ISM, and that fraction varies from $\sim 7\%$ to $\sim 90\%$ depending on the properties of the cloud. The work of \citet[][]{Geen2015} extends this analysis, focusing on the expansion of an ionization front in a turbulent, magnetized, self-gravitating molecular cloud. They focus on a series of very high resolution ($\leq 0.03$ pc) simulations of a $10^4 \Msun$ cloud embedded in a 27 pc box. They find that depending on the structure of the cloud and on the gas inflow rate, radiation can be completely trapped inside the cloud. Obviously, our cosmological scale simulation does not capture the inner structure of the cloud: the whole simulation volume of \citet{Geen2015} is described in our simulation by a few star-forming cells. Nevertheless, they provide an analytical framework to estimate the maximum expected size of an \hii region in a molecular cloud (see their eq.~15). This formalism relies on only a few parameters describing the density profile of molecular clouds, and once these parameters have been calibrated against high resolution, galactic scale simulations of molecular clouds, we could in principle apply this to our star-forming cells. This would yield a more realistic description of the escape of ionizing radiation from the stellar birth clouds, which are currently assumed to be mostly homogeneous. For instance, the simulation of a massive cloud presented in \citet{Howard2017} showed that environment of stellar clusters even inside the cloud plays an important role in the regulation of \fesc. Because in the simulations of \citet{Geen2015}, radiation leaks out of the clouds before the end of the main sequence evolution of the most massive stars, we would expect that the instantaneous \fesc increases, and that the time delay between the peaks of SFR and of \fesc would be reduced. However, developing this kind of framework would require a significant amount of work, since the study of \citet{Geen2015} only focuses on low-mass clouds.

\subsubsection{Stellar modelling}
\label{sec:stell-evol-modell}

While other sources of pre-SN feedback such as the inclusion of stellar winds or radiation pressure from infrared radiation have been suggested in the past, it is unclear that they might strongly affect the escape of ionizing photons. For instance, \citet{Geen2015a} suggested that the size of the gaseous shell formed by the main sequence stellar feedback is largely controlled by the radiative photoheating from the star, dominating over stellar winds. We note however that the study of \citet{Paardekooper2011} indicates that the pre-processing of the ISM in the vicinity of the ionizing sources by stellar feedback before SNe increases the ability of photons to escape the star forming cloud.

A more crucial assumption we made is the fact that all supernovae explode after 10 Myr. While this is the lifetime of a $\sim 15\ \Msun$ star, which would be a typical \textsc{SNii} progenitor, we should in principle sample the stellar lifetime properly, or at least use a variable time delay between the formation of a star particle and the release of energy due to the SN. Indeed, depending on the mass of the progenitor, SNe can occur as soon as 3 Myr, or as late as 40 Myr. However, we checked that changing the time delay in the simulation to 3 Myr has only a small impact on the escape of radiation: we compare in Fig.~\ref{fig:comparefb} the evolution of \fesc for the run with a delay of 10 Myr (in blue) and 3 Myr (in red), and the behaviour is qualitatively the same.

Finally, a note of caution regarding the stellar model we use is in order. We would like to draw the attention of the reader to the fact that we use a very small mass for our star particles, of only $\lesssim 150\ \Msun$. At these scales, the IMF cannot be fully sampled: \citet{Crowther2010} found, for instance, several stars with masses above 150 \Msun in the R136 star cluster in the Milky Way. We can obviously not capture the effect of these extremely massive stars with our setup. It would be necessary to stochastically sample the IMF, and use dedicated stellar evolution and SED models. However, as a result of our star formation and feedback schemes, the stars are very clustered spatially in our simulation, and there are groups of almost coeval star particles, somehow mitigating this issue.

On top of the various physical ingredients resulting in stellar feedback, the stellar population model itself might be of importance for a proper estimation of the ionizing budget. For instance, \citet{Topping2015} showed that stellar rotation increases the production of ionizing photons, over a longer period. \citet{Stanway2016} proposed that binary interactions might result in a similar enhancement of the ionizing luminosity. Recently, \citet{Ma2016} implemented this in their simulations, and find significantly higher escape fractions at all times. Runaway stars have also been advanced as a way to increase the photon production rate in high-$z$ galaxies \citet{Conroy2012}, but the work of \citet{Kimm2014} and \citet{Ma2015} ruled out any significant impact of such stars.

\subsection{Other limitations}
\label{sec:other-processes}

Apart from the small scale processes discussed in the previous section, our simulations make two supplementary simplifying assumptions: we include no external UV background, and we assume that in our galaxies, there is no additional feedback from a central, massive black hole.

Because we are mostly interested in the escape of radiation from the complex gas structure around low mass galaxies, we do not include any contribution from an external UV background. Recently, \citet{Simpson2013} studied the impact of various physical processes on the formation of a dwarf galaxies ($\Mvir \sim 10^9\ \Msun$), and found that the inclusion of a meta-galactic UV background has mainly the effect of removing diffuse gas around the halo, and that it is SN feedback that is mostly responsible of removing the dense gas that lies in the way of ionizing radiation. \citet{Paardekooper2015} showed that an external ionizing background results in the photoevaporation of gas in low mass haloes, helping to increase \fesc. At the same time, the star formation is strongly suppressed, therefore keeping the total contribution of very small haloes quite low. Overall, the main effect of the the external background is to remove the supply of fresh neutral gas, which would delay or block later star formation. Even so, the simulations of \citet{Rosdahl2012} constrain the global effect of the UV background to the gas at densities below $n_{\rm H} \lesssim 10^{-2}\ \mathrm{cm}^{-3}$, as denser gas is self-shielded. As a safety check, we reran the intermediate mass halo with an extra ionizing background, and found very little difference. This is because the haloes under study are too massive for star formation to be completely shut down.

Our simulations do not model the formation of massive black holes in the centre of our dwarf galaxies. This is aligned with most of the numerical studies focusing on the process of reionization by stellar sources. Indeed, while ubiquitously found in massive galaxies, it is only recently that \citet{Reines2013} discovered that massive black holes are not infrequent in low mass galaxies. Because high redshift galaxies are much more gas rich than their local counterparts, these black holes are likely to be more active. While we do not expect a strong direct contribution of these faint active galactic nuclei to the ionizing budget, the additional feedback due to the presence of a central massive black hole can in principle strongly affect the gas morphology, therefore affecting the escape of ionizing radiation.

\section{Conclusions}
\label{sec:summary}

Current constraints on the observed population of high redshift galaxies seem to indicate that low-mass galaxies play a significant role in the supply of ionizing photons needed to reionize the Universe by $z \sim 6$. Yet, their radiative efficiency, and in particular their escape fraction \fesc, is poorly constrained. To address this, we have performed a set of cosmological zoom simulations of high redshift dwarf galaxies, with very high spatial resolution ($\sim 7$ pc), and including a physically motivated subgrid description of star formation and supernova feedback. We have used the multifrequency, hydro-coupled radiative transfer module of \textsc{Ramses} to follow on the fly the emission of ionizing photons and their journey through the galactic halo to reach the intergalactic medium. While we do not have the statistical significance to assess directly the importance of the contribution of dwarf galaxies to the ionizing budget of reionization, we studied in great detail the processes affecting the evolution of the escape fraction, and our results are the following:

\begin{itemize}
\item The combination of spatially clustered star formation and efficient supernova feedback regulates the stellar mass of our simulated galaxies.
\item For galaxies with stellar masses of $\Mstar \sim 10^8-10^9\ \Msun$, the value of the instantaneous \fesc can vary by more than 6 orders of magnitude in $\sim 10$ Myr at a fixed \Mstar (Fig.~\ref{fig:fesc}). This variability can explain the large scatter in \fesc vs. \Mvir relation observed in simulations of cosmological volumes.
\item There is a delay between the peak of star formation and the peak of \fesc corresponding to the lifetime of massive stars (Fig.~\ref{fig:fesc_sfr_out}). This illustrate that the escape of ionizing radiation is mostly controlled by supernova feedback.
\item The escape of radiation from the halo is enabled by supernova feedback, resulting in a very anisotropic distribution of \fesc (Fig.~\ref{fig:direction}).
\item Without supernova feedback, all the radiation is absorbed locally within the stellar birth cloud (Fig.~\ref{fig:environment} and \ref{fig:r_combine}). When a massive star ends as a supernova, the mechanical energy removes gas from the neighbouring star forming clouds, thus clearing sight lines for radiation to escape.
\item While the possibility for a galaxy to leak ionizing radiation is mainly determined by cloud-scale properties, the gas distribution in the circumgalactic medium plays an important role in the determination of the exact value of \fesc at the virial radius, modulating the amount of ionizing radiation escaping the ISM.
\item The UV luminosity of a galaxy is only a very poor proxy for its ionizing emission: a UV-bright galaxy can be in a non-leaking phase (Fig.~\ref{fig:MUV_nesc}). Other diagnostics are needed to select promising Lyman-continuum emitters candidates.
\item Reciprocally, the mismatch between ionizing and non-ionizing luminosities means that even relatively deep UV limited surveys will systematically miss an important fraction of the sources of reionization.
\end{itemize}

This work depicts a scenario in which the escape fraction is a galaxy-scale quantity strongly tied to processes happening at the molecular cloud level.
Our study shows how stellar feedback is necessary to rearrange the gas in the halo to allow the release of radiation in the IGM. While photoionization on its own is not the dominant process in creating channels through which ionizing radiation can escape, it is still important to model it correctly. Indeed, previous work \citep[e.g.][especially their Fig.~3]{Rosdahl2015a} have shown that photoionization can contribute to smooth the ISM, thus altering the possible escape routes for ionizing photons.

The burstiness of the star formation history of such low mass galaxies might strongly impact their observability and inferred properties \citep[see e.g.][]{Dominguez2015}, the study of which is critical for the design of large surveys with the upcoming \emph{James Webb Space Telescope}. We plan to study this in further details in a future work.

\section*{Acknowledgements}

The authors wish to thank the anonymous referee for comments that helped improve this manuscript. We would like to thank Thibault Garel, Sam Geen, Taysun Kimm, Anne Verhamme and all the participants of the ORAGE project for useful discussion. We are very grateful to Leindert Boogaard for his careful bug-hunting.
MT and JB acknowledge support from the Lyon Institute of Origins under grant ANR-10-LABX-66, and from the ORAGE project from the Agence Nationale de la Recherche under grand ANR-14-CE33-0016-03. MT acknowledges funding from the European Research Council under the European Community's Seventh Framework Programme (FP7/2007-2013 Grant Agreement no. 614199, project `BLACK'), and the hospitality of the University of Oxford and New College through the award of a Balzan Fellowship. JR was funded by the European Research Council under the European Union's Seventh Framework Programme (FP7/2007-2013 Grant Agreement no. 278594, project `GasAroundGalaxies') and the Marie Curie Training Network CosmoComp (PITN-GA-2009-238356).
This work has made extensive use of the \textsc{PyMSES}\footnote{\label{fn:pymses}\url{https://bitbucket.org/dchapon/pymses/}} analysis package \citep{Guillet2013}.

%%%%%%%%%%%%%%%%%%%%%%%%%%%%%%%%%%%%%%%%%%%%%%%%%%



\bibliographystyle{mnras}
\bibliography{fescbib} % if your bibtex file is called example.bib

% \begin{thebibliography}{99}
% \end{thebibliography}

%%%%%%%%%%%%%%%%%%%%%%%%%%%%%%%%%%%%%%%%%%%%%%%%%%

% Don't change these lines
\bsp	% typesetting comment
\label{lastpage}
\end{document}